\documentclass{aa}  
\usepackage{comment}
\usepackage{csquotes}
\usepackage{longtable}
\usepackage{graphicx}
\usepackage{pdflscape}
\usepackage{color}
\usepackage{hyperref}
\usepackage[normalem]{ulem}
\usepackage{amsmath}	
\usepackage{amssymb}
\usepackage{txfonts}
\usepackage{mathtools}
\DeclarePairedDelimiter{\abs}{\lvert}{\rvert}
\def\degr{$^{\circ}$}
\newcommand{\dgr}{$^{\circ}$}

\newcommand{\kms}{km s$^{-1}$~}
\def\arcmin{^\prime}
\def\arcsec{^{\prime\prime}}

\begin{document}

   \title{Cloud Motion and magnetic fields: Four clouds in the Cepheus Flare region}

   \subtitle{}
   \titlerunning{Cloud Motion and magnetic fields: Four clouds in the Cepheus Flare region}
   \authorrunning{E Sharma et. al}

   \author{Ekta, Sharma,
          \inst{1,2}\thanks{ektasharma.astro@gmail.com},
          Maheswar G.\inst{1},
          Sami Dib \inst{3}
          }

   \institute{Indian Institute of Astrophysics (IIA), Sarjapur Road, Koramangala, Bangalore 560034, India
             \and
            Department of Physics and Astrophysics, University of Delhi, Delhi 110007, India
             \and
           Niels Bohr Institute, University of Copenhagen, {\O}ster voldgade 5-7, DK-1350, Copenhagen Denmark
           }

   \date{Received ; accepted }

 
  \abstract
   {The Cepheus Flare region consists of a group of dark cloud complexes that are currently active in star formation.}
   {The aim of this work is to estimate the motion of four clouds, L1147/1158, L1172/1174, L1228 and L1251 located at relatively high Galactic latitude ($b>14$\dgr) in the Cepheus Flare region. We study the relationship between the motion of the cloud with respect to the magnetic field and the clump orientations with respect to both the magnetic field and the motion.}
   {We estimated the motion of the molecular clouds using the proper motion and the distance estimates of the young stellar objects (YSOs) associated with them using the \textit{Gaia} EDR3 data. By assuming that the YSOs are associated with the clouds and share the same velocity, the projected direction of motion of the clouds are estimated. We measured the projected geometry of the magnetic field towards the direction of each cloud by combining the \textit{Planck} polarization measurements.}
   {We estimated a distance of 371$\pm$22 pc for L1228 and 340$\pm7$ pc for L1251 implying that all four complexes are located at almost the same distance. Assuming that both the clouds and YSOs are kinematically coupled, we estimated the projected direction of motion of the clouds using the proper motions of the YSOs. All the clouds in motion are making an offset of $\sim30$\degr~with respect to the ambient magnetic fields except in L1172/1174 where the offset is $\sim45$\degr. In L1147/1158, the starless clumps are found to be oriented predominantly parallel to the magnetic fields while prestellar clumps show random distribution. In L1172/1174, L1228 and L1251, the clumps are oriented randomly with respect to magnetic field. With respect to the motion of the clouds, there is a marginal trend that the starless clumps are oriented more parallel in L1147/1158 and L1172/1174. In L1228, the clumps' major axis are oriented more randomly. In L1251, we find a bimodal trend in case of starless clumps. We do not find any overall specific correlation between the core orientation and the global/local magnetic fields for the clouds in Cepheus. Also, we conclude that the local small-scale dynamics of the cloud with respect to the magnetic field direction could be responsible for the final orientation of the cores. }
   {}

   \keywords{Techniques: polarimetric -- Parallaxes, Proper motions -- Stars: formation -- ISM: clouds, magnetic field}

   \maketitle
%

%
\section{Introduction} \label{sec:intro}
The structure of the filamentary molecular clouds could be a consequence of the instabilities in the interstellar medium. The Parker instability can be thought as one of the factors responsible in the formation of filamentary molecular clouds \citep{2009MNRAS.397...14M,2020ApJ...891..157H}. Other possible mechanisms could be the converging flows \citep[e.g., ][]{1999A&A...351..309H, 1999ApJ...527..285B, 2006ApJ...643..245V, 2007arXiv0711.2417H, 2009ApJ...704..161I} driven by stellar feedback or turbulence. Stellar feedback processes like the expansion of HII regions \citep{1980ApJ...239..173B, 1995ApJ...441..702V, 1995ApJ...455..536P}, stellar-winds and supernova blast waves \citep{1987ApJ...317..190M, 1999ApJ...518..748G, 2000MNRAS.315..479D, 2001ApJ...551L..57D, 2005A&A...436..585D, 2006ApJ...638..797D,2009MNRAS.398.1201D,2011ApJ...743...25K, 2011ApJ...731...13N} can generate converging streams of gas that assemble to become molecular clouds, either in the Galactic plane or at relatively high Galactic latitudes. The energy input from supernovae and stellar winds can also eject material vertically up \citep[e.g., ][]{1990ARA&A..28...71S, 1993AIPC..278..499M, 1997ApJ...481..764B} which then flows back into the plane of the disk ($\text{v}_\text{gas} < \text{v}_\text{esc}$) through the diffuse magnetized interstellar medium (ISM). 


Magnetic fields may play a crucial role in the dynamics of cloud motion through the ISM. Using 2D numerical simulations \citet{1994ApJ...433..757M} showed that even a moderate level of magnetic field aligned parallel to the direction of the shock motion can help a cloud stabilize against disruptive instabilities \citep[see also ][]{1996ApJ...473..365J}. The magnetic field was found to have an even more dramatic impact when the motion of the cloud was considered transverse to the field lines \citep{1996ApJ...473..365J}. In this case, the field lines at the leading edge of the cloud get stretched creating a magnetic shield which quenches the disruptive Rayleigh-Taylor and Kelvin-Helmholtz instabilities and thus help the cloud survive longer. By varying the magnetic field orientation angle with respect to the cloud motion and also by varying the cloud-background density contrast and the cloud Mach number, \citet{1999ApJ...517..242M}, based on 2D numerical simulations showed that for large enough angles, the magnetic field tension can become significant in the dynamics of the motion of clouds having high density contrast and low Mach number. \citet{1999ApJ...527L.113G, 2000ApJ...543..775G}, using 3D numerical simulations of a moderately supersonic cloud motion through a transverse magnetic field, showed that the growth of dynamical instabilities get significantly enhanced due to the increase in the magnetic pressure at the leading edge of the cloud caused by the effective confinement of the magnetic field lines.

It is interesting to test these theoretical and numerical ideas on the Galactic molecular clouds. The Gould Belt is a distribution of stars and molecular clouds that forms a circular pattern in the sky having an inclination of $\sim20$\dgr with respect to the Galactic plane \citep{1879RNAO....1....1G}. The minimum and the maximum Galactic latitudes of the Gould Belt are toward the Orion and the Scorpio-Centaurus constellations respectively. The Cepheus Flare region is considered to be a part of the Gould Belt \citep[e.g.,][]{2009ApJS..185..198K}. This region is identified as having a complex of nebulae that extends 10\dgr-20\dgr~ out of the plane of the Galactic disk at a Galactic longitude of 110\dgr~ \citep{1934ApJ....79....8H, 1962ApJS....7....1L, 1987ApJ...315..104T, 1988ApJS...68..257C, 2002A&A...383..631D, 2005PASJ...57S...1D}. Five associations of dark clouds are found towards this region, namely,  L1148/1157, L1172/1174, L1228, L1241, and L1247/1251. There are signposts of current star formation in these cloud complexes \citep[e.g., ][]{2009ApJS..185..198K}.

Several shells and loops are identified in the direction of the Cepheus Flare. The Cepheus Flare Shell with its center at the Galactic coordinates of $l\sim120$\dgr~ and $b\sim17$\dgr, ~is considered to be an expanding supernova bubble and located at distance of $\sim300$ pc \citep{1989ApJ...347..231G, 2006MNRAS.369..867O}. Based on a study of the \ion{H}{i} distribution in the region of the Cepheus Flare, \citet{1969ApJ...156..493H} speculated the possibility of the presence of two sheets most likely representing an expanding or colliding system at a distance range of 300-500 pc. The presence of a giant radio continuum region \textit{Loop III} centered at $l=124\pm2$\dgr~, $b=+15.5\pm3$\dgr~ and extending across 65\dgr~\citep{1973A&A....24..143B}, suggests that it is possibly formed as a consequence of multiple supernova explosions. Also, the identification of an \ion{H}{i} shell by \citet{1981ApJ...248..119H} at $l=105$\dgr and $b=+17$\dgr~suggests that the ISM towards the Cepheus Flare region is far from being static.

\begin{figure}
\centering
  \includegraphics[width=8.9cm, height=8.5cm] {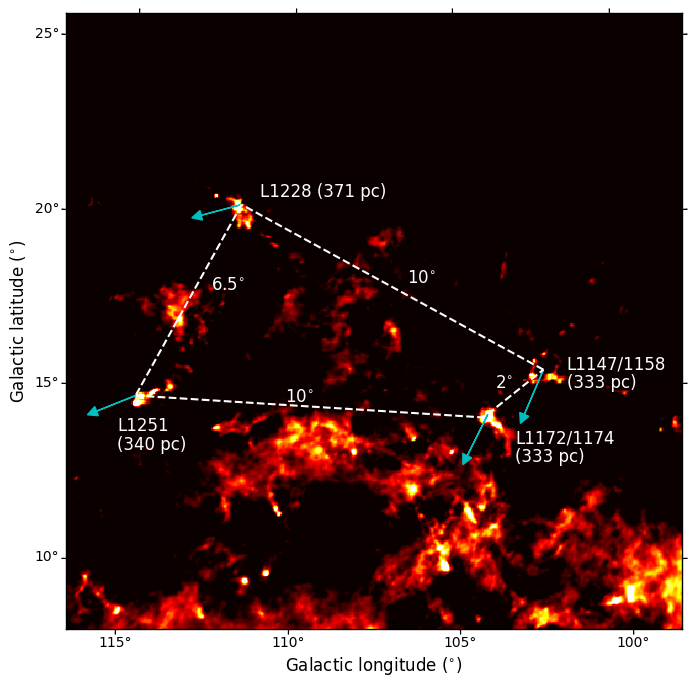}
  \caption{The 857 GHz \textit{Planck} image containing the Cepheus Flare region ($l\sim$ 100\degr~$-$ 116\degr and $b\sim$ 9\degr~$-$ 25\degr). The four cloud complexes, L1147/1158, L1172/1174, L1228 and L1251, studied here are identified and labelled. The arrows drawn in cyan show the direction of motion of the clouds in the sky plane inferred from the proper motion values (\textit{Gaia} EDR3) of the YSOs associated with the clouds. The median distances of the YSOs estimated using the \textit{Gaia} EDR3 in this study are also given.}\label{fig:cepheus_field}
\end{figure}

Based on a study on the kinematics of the gas in LDN 1157, a star forming cloud associated with the cloud complex LDN 1147-LDN 1158, \citet{2020A&A...639A.133S} showed that the southern boundary of the east-west segment was found to show a sinuous feature. Using the proper motion and parallax measurements of YSOs associated with L1147/1158 and L1172/1174 they suggested that the feature could be produced by the motion of the cloud through the ISM. In this paper, we studied the motion of two more cloud complexes, LDN 1228 \citep[hereafter L1228; ][]{1962ApJS....7....1L} and LDN 1251 \citep[hereafter L1251; ][]{1962ApJS....7....1L}. They are located in the direction of the Cepheus Flare region \citep{1997ApJS..110...21Y} and show signs of active star formation \citep{2009ApJS..185..451K}. Both these clouds are situated towards the east of L1147/1158 and L1172/1174 cloud complexes at an angular separation of $\sim10$\dgr~(Fig. \ref{fig:cepheus_field}). Using the parallax measurements of the YSOs associated with the clouds from the \textit{Gaia} EDR3, we find that both L1228 and L1251 are located at distances similar to L1147/1158 and L1172/1174 but show different radial velocities \citep{1989ApJS...71...89B, 1997ApJS..110...21Y, 2007ApJ...671.1748L}. Using the proper motion measurements of the YSO population identified in L1147/1158, L1172/1174, L1228 and L1251 and assuming that kinematically both YSOs and the clouds are coupled, we estimated the motion of the clouds on the plane of the sky. 

The plane of the sky component of the magnetic fields towards a molecular cloud are inferred from the observations made in optical \citep[e.g., ][]{1976AJ.....81..958V, 1990ApJ...359..363G, 2008A&A...486L..13A, 2013MNRAS.432.1502S, 2004ApJ...603..584P, 2015ApJ...807....5F, 2015A&A...573A..34S, 2016A&A...588A..45N}, near infrared \citep[e.g., ][]{1992ApJ...399..108G, 1995ASPC...73...45G, 2011ApJ...741...21C, 2010ApJ...716..299S, 2018ApJ...867...79C}, far-infrared \citep[e.g., ][]{2018ApJ...867...79C, 2019ApJ...872..187C, 2020NatAs.tmp..159P} and sub-mm wavelengths \citep{1998ApJ...502L..75R, 2004A&A...424..571B, 2007SPIE.6678E..0DV, 2010ApJS..186..406D, 2016A&A...586A.138P}. Since the interstellar dust grains align their long axis perpendicular to the magnetic field, the background star light polarization is parallel to the local magnetic field. In far-infrared and sub-mm wavelengths, thermal dust emission is found to be linearly polarized with the polarization position angles lying perpendicular to the ambient magnetic field orientations \citep{2004A&A...424..571B, 2007SPIE.6678E..0DV}. This provides information on the geometry of the projected magnetic field lines. The magnetic field morphology traced by the polarization measurements made with the \textit{Planck} are used not only to infer the Galactic magnetic field geometry, but also to place new constraints on the dust grain properties \citep{2015A&A...576A.106P, 2016A&A...586A.135P, 2016A&A...586A.138P, 2016A&A...586A.141P}. In this study, the plane of the sky component of the magnetic field geometry was inferred for all the four clouds using the \textit{Planck} polarization measurements.

Having information on the motion of the cloud and the magnetic field geometry on the plane of the sky, we studied various relations such as (a) the relative orientation between the projected magnetic field and the direction of motion of the clouds (b) the relative orientation between the cores' major axis and the orientation of the magnetic fields (both within the cloud and in the inter-cloud region) and (c) relative orientation between the cores' major axis and the direction of motion of the clouds. This paper is organized in the following manner. We begin with a description on the data used in section \ref{sec:data} followed by a discussion on the YSO population in each cloud, the estimation of the distances of the clouds, proper motion values of the YSOs in section \ref{sec:res_dis}. The motion of the clouds are discussed with respect to the orientation of the projected magnetic fields and the orientation of the clumps identified in the clouds with respect to both the magnetic fields and the direction of motion are discussed further. We finally conclude our paper with a summary of the results in section \ref{sec:con}.

\section{The Data}\label{sec:data}

\subsection{The Gaia EDR3}
The \textit{Gaia} Early Data Release 3 (EDR3) provides five-parameter astrometric solution, namely, positions on the sky ($\alpha$, $\delta$), parallaxes, and proper motions for more than 1.5 billion sources \citep{2021A&A...649A...1G}.  The limiting magnitude is $\sim21$ in G-band. Uncertainties in the parallax values are in the range of $\sim$ 0.02-0.03 milliarcsecond for sources with G $\leq$ 15, and around 0.07 mas for sources with G $\sim$ 17 mag. For the sources at the fainter end, with G$\sim$20, the uncertainty in parallax is of the order of 0.5 mas \citep{2021A&A...649A...2L}. The standard uncertainties in \text{Gaia} EDR3 as compared to DR2 on an average has improved roughly by a factor of 0.8 for the positions and parallaxes, and 0.5 for the values of proper motions. The uncertainties in the corresponding proper motion values are up to 0.01-0.02 mas/yr for G $\leq$ 15 mag, 0.05 mas/yr for G = 17 mag and 0.4 mas/yr for G = 20 mag. The conversion from parallax to distance is known to become non-trivial when the observed parallax is small compared to its uncertainty especially in cases where $\sigma_{\varpi}$/$\varpi\gtrsim20\%$ \citep{2015PASP..127..994B}. As a result, by adopting an exponentially decreasing space density prior in distance, \cite{2015PASP..127..994B} estimated distances of about 1.47 billion sources using the \textit{Gaia} parallax measurements. The distances to the YSOs studied in this work are obtained from the catalogue provided by \citet{2021AJ....161..147B} while the proper motion values are obtained from the \textit{Gaia} EDR3 \citep{2021A&A...649A...1G} catalog. We used only those values for which the ratio, m/$\Delta$m (where m represents either the distances or the proper motions and $\Delta$m is the corresponding uncertainties in two quantities), is greater than or equal to 3.  

\subsection{The Planck 353 GHz polarization measurements}\label{sec:plk_pol}

The 353 GHz (850 $\mu$m) channel is the highest-frequency polarization-sensitive channel of the Planck. We constructed the geometry of the magnetic fields in the vicinity of L1147/1158, L1172/1174, L1228 and L1251 based on these data. We selected images containing the clouds and smoothed them down to the 8$\arcmin$ resolution to obtain a good signal-to-noise (SNR) ratio. As mentioned before, the dust emission is linearly polarized with the electric vector normal to the sky-projected magnetic field, therefore the polarization position angles were rotated by 90\degr~ to infer the projected magnetic field.

\subsection{Identification of clumps from the Herschel column density maps}\label{ide_clump}
In our analysis, we used the sample of cores identified by \cite{2020ApJ...904..172D} using \textit{Getsources} on high resolution column density images of four complexes. The method  \textit{Getsources} is a multi-scale method to filter out emission from monochromatic images. Out of all the sources, we have considered only those sources where the aspect ratio is less than 0.8. 

To check the dependency of our analysis on any core-extraction method, we also identified clumps in the column density maps of the four clouds using the \textit{Astrodendro} algorithm, which is a Python package to identify and characterize hierarchical structures in images and data cube. The dust column density maps of the four clouds L1147/1158, L1172/1174, L1228 and L1251 were obtained from the Herschel Gould Belt Survey Archive (\url{http://www.herschel.fr/cea/gouldbelt}). A dendrogram is a tree diagram which identifies the hierarchical structures of 2-dimensional and 3-dimensional datasets. The structures start from the local maximum with the volumes getting bigger when it merges with the surroundings with lower flux densities and stop identifying structures when they meet neighbouring structures \citep{2008ApJ...679.1338R}. The dendrogram method identifies emission features at successive isocontours in emission maps which are called \textit{leaves} and find the intensity values at which they merge with the neighbouring structures (branches and trunks). In order to extract the structures, three parameters need to be defined : min\textunderscore{value}, min\textunderscore{delta} and min\textunderscore{npix}. The parameter, min\textunderscore{value} is an emission threshold above which all the structures are identified and min\textunderscore{delta} is contour interval which decides the boundary between the distinct structures. The initial threshold and the contour step size are selected as a multiple of the $\sigma$, the rms of the intensity map. In our analysis, we are interested in the denser regions which are identified as leaves. 

In each cloud, the minimum threshold was selected as a multiple of the background column density values, $\sim$ 0.5-65 $\times$ 10$^{21}$ cm$^{-2}$. We considered emission free regions well away from the cloud emission to find out the background column density. The other parameter, min$\textunderscore$delta decides whether the main structure is to be identified as independent entity or merged with the main structure. After trying a set of choices, we used twice of rms of the emission map to minimize picking up of noise structure. In order to identify the real structures, we used the condition on the size of the clump as that the area within each ellipse should be higher than the area within 30 pixels for the high-resolution image. Second, we excluded the clumps where the area of each source is smaller than the area subtended by a beam. Here, we used the Herschel column density map for the source extraction with the beam size $\sim$ 18$\arcsec$. In addition to that, to consider only the elongated clumps, we used only those clumps where the aspect ratio is less than 0.8 \citep{2020MNRAS.494.1971C}. Since we are interested in the orientation of cores embedded within the large clouds, we removed those clumps from our sample that are lying on the edges of the dust column density maps. 

\section{Results and Discussion}\label{sec:res_dis}

\subsection{Population of YSOs, distance and proper motion}\label{sec:ysos}
\begin{figure}
\centering
  \includegraphics[width=8.7cm, height=8cm]{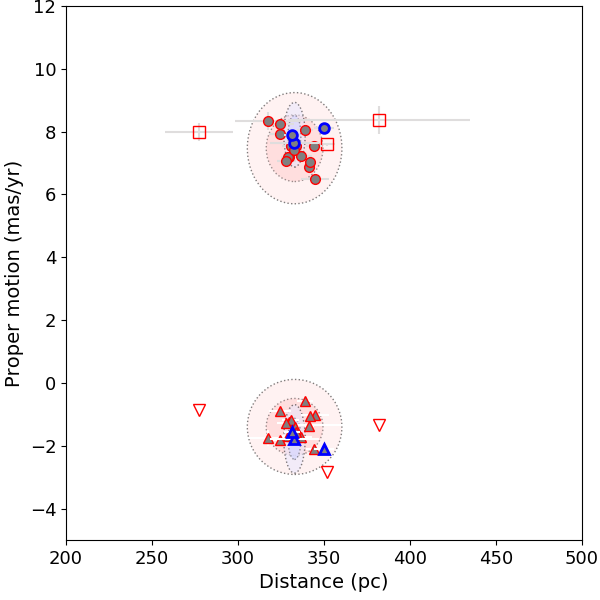} 
  \caption{Proper motion vs. distance plot for stars towards L1147/1158 (blue symbols) and L1172/1174 (red symbols). The $\mu_{\alpha\star}$ and $\mu_{\delta}$ are represented using circles and triangles respectively. Sources listed in Table \ref{tab:yso_dist_pm} for L1147/1158 and L1172/1174 are shown using blue filled triangles/circles and red filled triangles/circles, respectively. The darker and lighter shaded ellipses are drawn using 3 and 5 times the MAD values of the distance and the proper motions respectively. The sources falling outside of the 5 times the MAD ellipse are identified using squares ($\mu_{\alpha\star}$) and inverted triangles ($\mu_{\delta}$).}\label{fig:pm_dist}
\end{figure}

The star forming regions in the Cepheus Flare was extensively studied by \citet{2009ApJS..185..198K}, \citet{2009ApJS..185..451K}, and \citet{2013MNRAS.429..954Y}. The YSOs used in this study are taken from the above three works. We obtained the \textit{Gaia} EDR3 data for the three YSO candidates that are located in the direction of L1147/1158 cloud complex \citep{2009ApJS..185..198K}. These sources are listed in Table \ref{tab:yso_dist_pm}. Columns 1-5 give source names, right ascension, declination, Galactic longitude and latitude. Columns 6-10 give the distances obtained from the \citet{2021AJ....161..147B} catalog, $\mu_{\alpha\star}$ ($=\mu_{\alpha}\text{cos}\delta$) and $\mu_{\delta}$ along with the uncertainties obtained from the \textit{Gaia} EDR3 \citep{2021A&A...649A...1G} catalog. The distance and proper motion values of these three sources associated with L1147/1158 cloud complex are shown in Fig. \ref{fig:pm_dist} using open circles ($\mu_{\alpha\star}$) and triangles ($\mu_{\delta}$) in blue. The median and the standard deviation values of the distances and $\mu_{\alpha\star}$ and $\mu_{\delta}$ are found to be 333$\pm$1 pc, 7.764$\pm$0.137 mas/yr and -1.672$\pm$0.108 mas/yr respectively. These values are given in columns 2, 3 and 4 of Table \ref{tab:yso_dist_pm_mean}.

Recently, \citet{2020MNRAS.494.5851S} made a study of the YSO candidates in the direction of L1172/1174 based on the \textit{Gaia} DR2 data. They obtained the \textit{Gaia} DR2 data for a total of 19 known YSOs compiled from \citet{2009ApJS..185..198K}, \citet{2009ApJS..185..451K} and \citet{2013MNRAS.429..954Y}. The filled circles and triangles in red show results for L1172/1174 in Fig. \ref{fig:pm_dist}. The median and median absolute deviation (MAD), more robust against outliers, for the distances, $\mu_{\alpha\star}$ and $\mu_{\delta}$ are 333$\pm$6 pc, 7.473$\pm$0.353 mas/yr and  -1.400$\pm$0.302 mas/yr respectively. In Fig. \ref{fig:pm_dist}, we also draw ellipses with 3$\times$MAD (darker shade) and 5$\times$MAD (lighter shade) for L1172/1174. Of the 19 sources found towards L1172/1174, three sources are found to fall outside the ellipses drawn with 5$\times$MAD. These sources are shown using open squares ($\mu_{\alpha\star}$) and inverted triangles ($\mu_{\delta}$) respectively. Only the sources that fall within the limit of 5$\times$MAD ellipses are considered for our study because these sources with the significant values of proper motion and distance are considered to be a member of the particular complex and the remaining ones are considered as outliers. The three sources identified in the direction of L1147/1158 are also found to fall within the 5$\times$MAD ellipses obtained for L1172/1174 cloud. These sources are listed in Table \ref{tab:yso_dist_pm}. 

\begin{figure}
\centering
  \includegraphics[width=8.7cm, height=8cm]{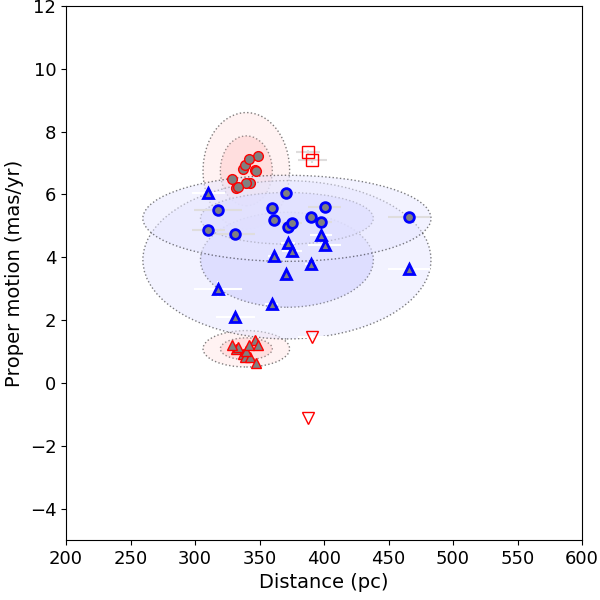}
  \caption{Proper motion vs. distance plot for stars towards L1228 (blue) and L1251 (red). The $\mu_{\alpha\star}$ and $\mu_{\delta}$ are represented using circles and triangles respectively. Sources listed in Table \ref{tab:yso_dist_pm} for L1228 and L1251 are shown using blue filled triangles/circles and red filled triangles/circles, respectively. The darker and lighter shaded ellipses are drawn using 3 and 5 times the MAD values of the distance and the proper motions respectively. The sources falling outside of the 5 times the MAD ellipse are identified using squares ($\mu_{\alpha\star}$) and inverted triangles ($\mu_{\delta}$).}\label{fig:pm_dist_1}
\end{figure}
\begin{figure}
\centering
  \includegraphics[width=8.7cm, height=8cm]{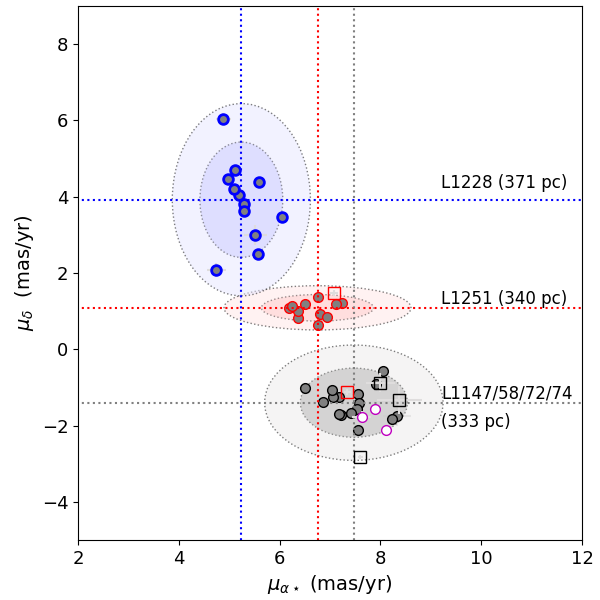}
  \caption{Proper motion values of sources associated with L1147/1158, L1172/1174, L1228 and L1251. The darker and lighter shaded ellipses are drawn using 3 and 5 times the median absolute deviation values of the proper motions, respectively. The sources lying outside of the 5 times the MAD ellipse are identified using squares symbols.}\label{fig:pm_pm}
\end{figure}

The YSO candidates towards L1228 and L1251 are obtained from the source list provided by \citet{2009ApJS..185..451K}. We obtained \textit{Gaia} EDR3 data for 12 and 13 sources towards L1228 and L1251 respectively. The results are shown using the filled circles and triangles in blue (L1228) and red (L1251) respectively. The mean and the MAD values of the distance and the proper motions, $\mu_{\alpha\star}$ and $\mu_{\delta}$ obtained for L1147/1158, L1172/1174, L1228 and L1251 complexes are given in columns 2, 3 and 4 of Table \ref{tab:yso_dist_pm_mean}. Two sources belonging to L1251 are found to fall outside the ellipses drawn with 5$\times$MAD which are identified using open squares and inverted triangles in red (L1251). The sources that are falling within the 5$\times$MAD ellipses towards L1228 (12) and L1251 (11) respectively are listed in Table \ref{tab:yso_dist_pm}. 


A clustering of the sources associated with each cloud complex is quite evident. It is also apparent that both L1147/1158 and L1172/1174 are located at similar distances from us even though they are $\sim$2\degr~apart in the sky. This implies that they are spatially $\sim10$ pc apart. L1228 which is located at an angular distance of $\sim$10\degr~away from both L1147/1158 and L1172/1174 is slightly further away at $\sim$371 pc from us. This translates to a spatial separation of $\sim$60 pc. The errors in the distances for two sources in L1228 from Gaia EDR3 are higher as compared to Gaia DR2 measurements.
L1251 which is spatially at a separation of $\sim$6\degr~away from L1228 is located at a distance of $\sim$340 pc. This cloud lies close to the complexes, L1147/1158 and L1172/1174. L1251 is also at a spatial separation of $\sim$10\degr~away from both L1147/1158 and L1172/1174. In Fig. \ref{fig:pm_pm}, we show the $\mu_{\alpha\star}$ and $\mu_{\delta}$ values for L1147/1158 and L1172/1174 (magenta + grey), L1228 (blue) and L1251 (red). The median values are identified using dotted lines and the ellipses are drawn using 3$\times$MAD (darker shade) and 5$\times$MAD (lighter shade). The square symbols show the sources that fall outside 5$\times$MAD ellipses in Fig. \ref{fig:pm_dist} and hence are not considered in this study. In proper motion domain also, all but one in L1148/57/72/74 are falling within the 5$\times$MAD ellipses in Fig. \ref{fig:pm_pm}. There is a clear clustering of sources implying that the YSO candidates associated with the clouds are moving coherently in space.

In earlier studies, L1228 was considered to be a cloud lying closer to us at a distance of $\sim200$ pc \citep[e.g., ][]{2008hsf1.book..136K}. Kinematically also this cloud differs from the rest of the clouds belonging to the Cepheus Flare region. At least three centers of star formation were identified towards L1228 \citep[e.g., ][]{2008hsf1.book..136K}. 
Recently, \citet{2020A&A...633A..51Z} estimated distances to the molecular clouds by inferring the distance and extinction to stars lying projected on the clouds using optical and near-infrared photometry and the \textit{Gaia} DR2 parallax measurements. Based on their results, the distances provided for the regions enclosed within the Galactic coordinates $l\sim102$\degr-115\degr~and $b\sim14$\degr-21\degr~ are found to be in the range of $\sim330$ pc to $\sim380$ pc. This is quite consistent with the distances estimated for the four molecular clouds presented here using the YSO candidates associated with the cloud which is the most direct method of estimating distances to molecular clouds.

Thanks to the \textit{Gaia} mission, now we have parallax and proper motion measurements for over billion stars with unprecedented accuracy. Though the high-resolution spectroscopy of stars can provide the measurements of radial velocity, it is still not possible to get the measurements made for YSOs as they are typically fainter in magnitudes. On the other hand, it is not possible to measure the proper motion of the molecular clouds as they are not point sources. Therefore by combining the proper motion measurements of the YSOs associated with it and by assuming that both the YSOs and the clouds share the same motion as these clouds are the birth places of the YSOs, we can determine the projected direction of motion of a molecular cloud in the plane-of -sky. The proper motions of the sources measured by the \textit{Gaia} are in the equatorial system of coordinates. To understand the motion of objects in the Galaxy, we need to transform the proper motion values from the equatorial to the Galactic coordinate system $\mu_{l\star}=\mu_{l}$ cos$b$ and $\mu_{b}$. We transformed the proper motion values using the expression \citep{2013arXiv1306.2945P},
\begin{equation}
\left[\begin{array}{c}\mu_{l\star}\\\mu_{b}\end{array}\right]=\frac{1}{cosb}\begin{bmatrix}\text{C}_{1} & \text{C}_{2}\\-\text{C}_{2} & \text{C}_{1}\end{bmatrix}\left[\begin{array}{c}\mu_{\alpha\star}\\\mu_{\delta}\end{array}\right]
\end{equation}
where the term $cosb$ = $\sqrt{\text{C}_{1}^{2} + \text{C}_{2}^{2}}$ and the coefficients C$_{1}$ and C$_{2}$ are given as,
\begin{equation}
\begin{array}{l@{}l}
\text{C}_{1} = sin\delta_{G}~cos\delta - cos\delta_{G}~sin\delta~cos(\alpha-\alpha_G) \\
\text{C}_{2} = cos\delta_{G}~sin(\alpha-\alpha_G)
\end{array}
\end{equation}
The equatorial coordinates ($\alpha_{G}$, $\delta_{G}$) of the North Galactic Pole are taken as 192\degr.85948 and 27\degr.12825, respectively \citep{2013arXiv1306.2945P}. The mean values of $\mu_{l\star}$ and $\mu_{b}$ calculated for the YSO candidates found towards the four cloud complexes are given in columns 5 and 6 of Table \ref{tab:yso_dist_pm_mean}. If we assume that the cloud and the YSO candidates are expected to share similar kinematics as a result of them being born inside the cloud, then the arrows should also represent the motion of the clouds on the plane-of-the-sky. The presence of a reflection nebulosity around a number of these YSO candidates provides evidence of their clear association with the cloud. The plane of the sky component of the direction of motion ($\theta^{motion}_{pos}$) calculated using the mean values of $\mu_{l\star}$ and $\mu_{b}$ are given in the column 7 of Table \ref{tab:yso_dist_pm_mean}. The arrows drawn in black in Fig. \ref{fig:pol-plk-gal} show the sense of the motion of the sources on the sky plane with respect to the Galactic north increasing to the east.  

We also computed the scale height of the complexes above the Galactic plane which is 87 pc for L1147/1158, 79 pc for L1172/1174, 121 pc for L1228 and 83 pc for L1251. All  four clouds are lying at a height in the range of 83 - 121 pc with the furthest being L1228 which is located at a height of $\sim121$ pc above the Galactic mid-plane. These displacements above the Galactic plane are larger compared with the rms Z-dispersion of the clouds located within 1 kpc \citep[$\sim80$ pc; ][]{1987ApJ...322..706D, 2005ARA&A..43..337C}.  

\begin{table*}
\begin{small}
\caption{\textit{Gaia} DR3 results of YSOs associated with the L1147/1158, L1172/1174, L1228 and L1251 complexes.}\label{tab:yso_dist_pm}
\renewcommand{\arraystretch}{1.4}
\begin{tabular}{lccccccccc} \hline
Source Name				&RA             &Dec           & $l$& $b$      & D   &$\mu_{\alpha}$ &$\Delta\mu_{\alpha}$ & $\mu_{\delta}$ &$\Delta\mu_{\delta}$\\
                        & ($^{\circ}$)  & ($^{\circ}$) & ($^{\circ}$)  & ($^{\circ}$) &(pc)& (mas/yr) & (mas/yr)& (mas/yr) & (mas/yr)\\
     (1)&(2)&(3)&(4)&(5)&(6)&(7)&(8)&(9)&(10)\\\hline\hline
     \multicolumn{10}{c}{\bf L1147/1158}\\
2MASS J20361165+6757093	&309.048672&	$+$67.952608& 102.4221&	$+$15.9738&	333$^{15}_{-15}$& 7.627&  0.183&   -1.781&  0.155 \\
IRAS 20359+6745			&309.082855&	$+$67.942131& 102.4205&	$+$15.9573&	332$^{4}_{-5}$&	7.902&  0.055&   -1.563& 0.045 \\                                                                                  
PV Cep			&311.474902&	$+$67.960735 & 102.9697&	$+$15.2315&	350$^{6}_{-5}$&	 8.108& 0.057& -2.108&  0.054 \\\hline
\multicolumn{10}{c}{\bf L1172/1174}\\
FT Cep			&314.845315&	$+$68.245467&	103.9926&	$+$14.4053&	329$^{2}_{-2}$&		7.184 &	0.024 & 	-1.239 &	0.023 \\ 
2MASS J21002024+6808268	&315.084447&	$+$68.140772&	103.9661&	$+$14.2704&	331$^{5}_{-4}$&		7.557 &	0.064 &	-1.181 &	0.064 \\
2MASS J21005550+6811273	&315.231481&	$+$68.190885&	104.0418&	$+$14.2596&	337$^{6}_{-7}$&		7.214 &	0.068 & 	-1.719 &	0.080 \\
NGC 7023 RS 2		&315.359984&	$+$68.177338&	104.0621&	$+$14.2140&	333$^{2}_{-2}$&		7.569 &	0.026 & 	-1.416 &	0.024 \\
NGC 7023 RS 2B		&315.362884&	$+$68.177214&	104.0627&	$+$14.2131&	345$^{8}_{-8}$&		6.508 &	0.078 & 	-1.016 &	0.074 \\
LkH$\alpha$ 425		&315.400352&	$+$68.139576&	104.0418&	$+$14.1785&	329$^{4}_{-4}$&		7.182 &	0.055 & 	-1.687 &	0.051 \\
HD 200775		&315.403923&	$+$68.163263&	104.0616&	$+$14.1924&	352$^{5}_{-5}$&		7.597 &	0.049 & 	-2.821 &	0.046 \\
NGC 7023 RS 5		&315.427117&	$+$68.215960&	104.1093&	$+$14.2191&	339$^{4}_{-5}$&		8.043 &	0.056 & 	-0.578 &	0.053 \\
FU Cep			&315.444875&	$+$68.145894&	104.0577&	$+$14.1696&	334$^{2}_{-3}$&		7.528 &	0.024 & 	-1.556 &	0.024 \\
FV Cep			&315.558650&	$+$68.233141&	104.1549&	$+$14.1923&	317$^{21}_{-17}$&		8.333 &	0.281 & 	-1.754 &	0.262 \\
LkH$\alpha$ 428 N		&315.617758&	$+$68.058287&	104.0301&	$+$14.0642&	341$^{3}_{-2}$&		6.871 &	0.034 & 	-1.385 &	0.029 \\
FW Cep			&315.637634&	$+$68.124746&	104.0878&	$+$14.1008&	342$^{1}_{-2}$&		7.031 &	0.019 & 	-1.061 &	0.016 \\
NGC 7023 RS 10		&315.747855&	$+$68.108939&	104.1022&	$+$14.0590&	333$^{5}_{-4}$&		7.419 &	0.050 & 	-1.668 &	0.046 \\
EH Cep			&315.851719&	$+$67.985134&	104.0295&	$+$13.9501&	324$^{8}_{-8}$&		8.229 &	0.093 & 	-1.821 &0.091 \\
2MASS J21035938+6749296	&315.997585&	$+$67.824847&	103.9389&	$+$13.8053&	344$^{5}_{-5}$&		7.552 &	0.048 & 	-2.105 &	0.046 \\\hline
\multicolumn{10}{c}{\bf L1228}\\
2MASS J20584668+7740256&   314.694891&  $+$77.673875&   111.7897&  +20.1974& 397$^{9}_{-8}$   & 5.109&	0.099 	&4.712	&0.092\\
2MASS J20590373+7823088&   314.765753&  +78.385837&   112.4115&  +20.6066&   331$^{16}_{-13}$   & 4.741&	0.191& 	2.085&	0.189\\
2MASS J21005285+7703149&   315.220237&  $+$77.054169&   111.3338&  +19.7352& 400$^{14}_{-11}$   & 5.586&	0.092 &	4.384&	0.093 \\
2MASS J21012919+7702373&   315.370747&  $+$77.043700&   111.3465&  +19.7019& 390$^{5}_{-4}$     & 5.297&	0.032 &	3.801&	0.032\\
2MASS J21012919+7702373&   315.372805&  $+$77.043769&   111.3468&  +19.7016& 361$^{5}_{-4}$     & 5.185&	0.039 &	4.041&	0.037\\
2MASS J21013097+7701536&   315.379197&  +77.031566&   111.3374&  +19.6931& 310$^{16}_{-10}$    & 4.883	&0.144 &	6.045&	0.146 \\
2MASS J21013267+7701176&   315.385687&  $+$77.021545&   111.3299&  +19.6859& 372$^{5}_{-6}$     & 4.969&	0.039& 	4.466&	0.040 \\
2MASS J21014960+7705479&   315.456720&  $+$77.096668&   111.3299&  +19.6859& 375$^{7}_{-8}$   & 5.097&	0.067 &	4.193&	0.068\\
2MASS J21020488+7657184&   315.520565&  $+$76.955170&   111.2933&  +19.6216& 371$^{1}_{-2}$     & 6.039&	0.013 &	3.474&	0.012\\
$^{\dagger}$2MASS J21030242+7626538&   315.759650&  $+$76.448351&   110.9023&  +19.2695& 318$^{19}_{-17}$   & 5.517&	0.226& 	2.994&	0.257\\
2MASS J21030242+7626538&   315.761082&  $+$76.448379&   110.9025&  +19.2692& 359$^{5}_{-5}$     & 5.580&	0.046 &	2.498&	0.069\\
2MASS J21055189+7722189&   316.466412&  $+$77.371980&   111.7800&  +19.7100& 466$^{17}_{-15}$     &	5.297	&0.083 &	3.618	&0.069\\\hline
\multicolumn{10}{c}{\bf L1251}\\
2MASS J22351668+7518471  & 338.819607&      $+$75.313082&	114.5620&      $+$14.7206&	 347$^{3}_{-2}$	&6.770&	0.025& 	1.375&	0.024\\
2MASS J22352542+7517562 &	338.855984&      $+$75.298960&	114.5626&      $+$14.7037&	 337$^{5}_{-5}$	&6.797&	0.050& 	0.918&	0.044\\
2MASS J22352722+7518019 &	338.863513&      $+$75.300554&	114.5652&      $+$14.7041&	 339$^{5}_{-5}$	&6.940&	0.055& 	0.841&	0.056\\
2MASS J22381872+7511538 &	339.578072&      $+$75.198277&	114.6731&      $+$14.5234&	 349$^{4}_{-4}$	&7.239&	0.040& 	1.203&	0.044\\
2MASS J22374953+7504065 &	339.456610&      $+$75.068466&	114.5776&      $+$14.4271&	 343$^{1}_{-2}$	&6.373&	0.016& 	0.819&	0.019\\
2MASS J22392717+7510284 &	339.863479&      $+$75.174575&	114.7260&      $+$14.4662&	 340$^{3}_{-3}$	&6.365&	0.026& 	1.012&	0.024\\
2MASS J22381522+7507204 &	339.563569&      $+$75.122351&	114.6302&      $+$14.4597&	 348$^{7}_{-5}$	&6.752&	0.074& 	0.643&	0.075\\
2MASS J22382962+7514266 &	339.623507&      $+$75.240739&	114.7056&      $+$14.5542&	 331$^{9}_{-9}$	&6.195&	0.087& 	1.087&	0.095\\
2MASS J22384046+7508413 &	339.668708&      $+$75.144805&	114.6660&      $+$14.4655&	 333$^{8}_{-9}$	&6.250&	0.110& 	1.133&	0.123\\
2MASS J22391466+7507162 &	339.811230&      $+$75.121196&	114.6864&      $+$14.4267&	 342$^{6}_{-5}$	&7.123&	0.054& 	1.196&	0.054\\
2MASS J22410470+7510496 &	340.269829&      $+$75.180475&	114.8221&      $+$14.4196&	 329$^{5}_{-5}$	&6.500&	0.052& 	1.197&	0.052\\
\hline
\end{tabular}

$^{\dagger}$There are two sources within 1$^{\prime\prime}$ search radius.
\renewcommand{\arraystretch}{1}
\end{small}
\end{table*}

\begin{table*}
\begin{small}
\caption{\textit{Gaia} DR3 distances, proper motion and direction of magnetic field for L1147/1158, L1172/1174, L1228 and L1251 complexes.}\label{tab:yso_dist_pm_mean}
\centering
\renewcommand{\arraystretch}{1.43}
\begin{tabular}{lccccccc} \hline
Cloud 	& D  &$\mu_{\alpha}$  & $\mu_{\delta}$ & $\mu_{l\star}$ & $\mu_{b}$  &$\theta^{motion}_{pos}$   & $\theta^{cloud}_{Bpos}$, $\theta^{ICMF}_{Bpos}$ \\ 
Complex	& (pc)	    & (mas/yr)		 & (mas/yr)	       & (mas/yr)	    & (mas/yr)  &(\dgr) &(\dgr)\\ \hline
(1)      & (2)       &(3)             &(4)              &(5)             & (6)       & (7) & (8) \\ \hline
L1147/1158      &333$\pm$1	&7.764$\pm$0.137  &-1.672$\pm$0.108 & 2.974  & -7.520 &158 & 183$\pm$7, 186$\pm$6  		\\
L1172/1174      &333$\pm$6	&7.473$\pm$0.353  &-1.400$\pm$0.302 & 3.646 & -6.645 &151 & 208$\pm$11, 196$\pm$23	\\
L1228           &371$\pm$22 &5.241$\pm$0.274  & 3.921$\pm$0.504 & 6.291   &-1.962 &107& 84$\pm$7, 88$\pm$10  \\
L1251			&340$\pm$7 & 6.752$\pm$0.371 & 1.087$\pm$ 0.116 & 6.273 & -2.474 & 112 &  79$\pm$6, 81$\pm$6 \\ \hline
\end{tabular}

\renewcommand{\arraystretch}{1.3}
\end{small}
\end{table*}

\subsection{Motion of the cloud complexes with respect to the magnetic field}\label{sec:mag}

\begin{figure*}
     \includegraphics[width=8.93cm, height=6cm]{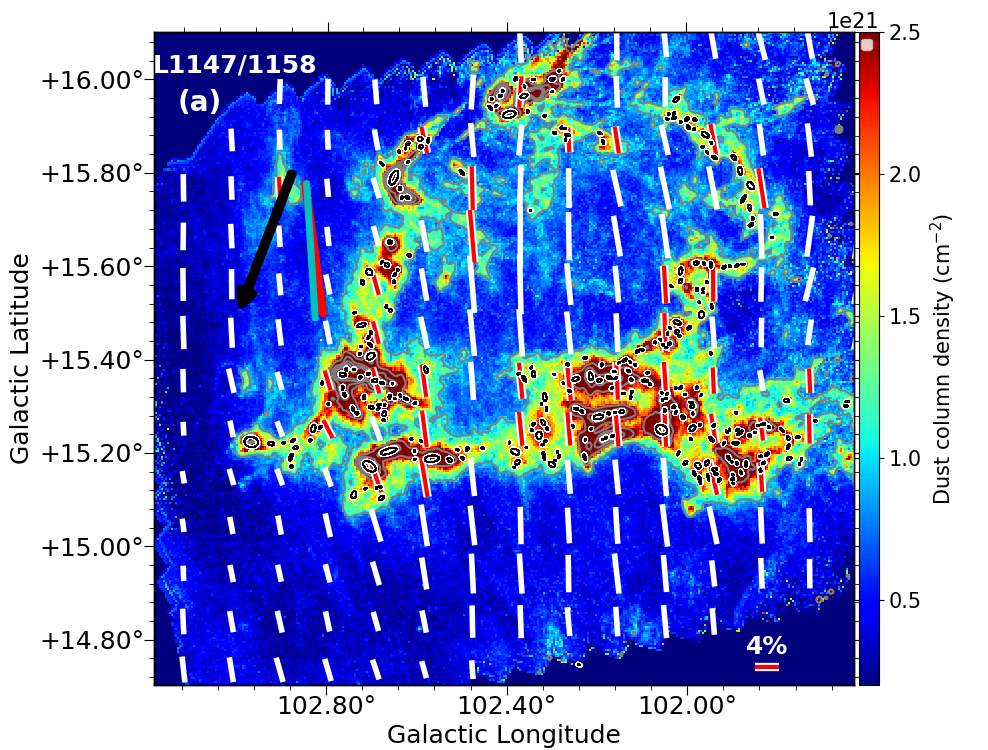}
   \includegraphics[width=8.9cm, height=6cm]{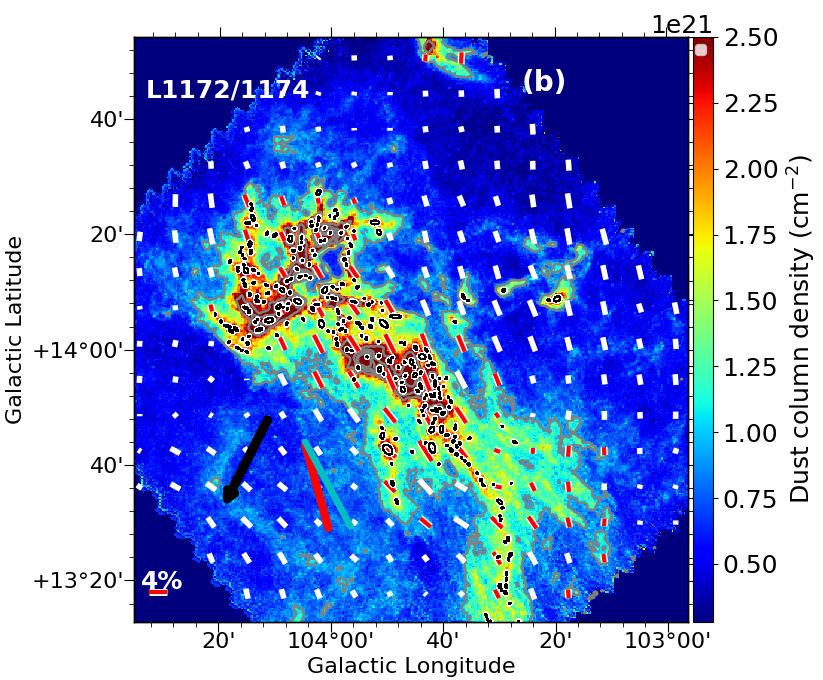}
   \includegraphics[width=8.95cm, height=6cm]{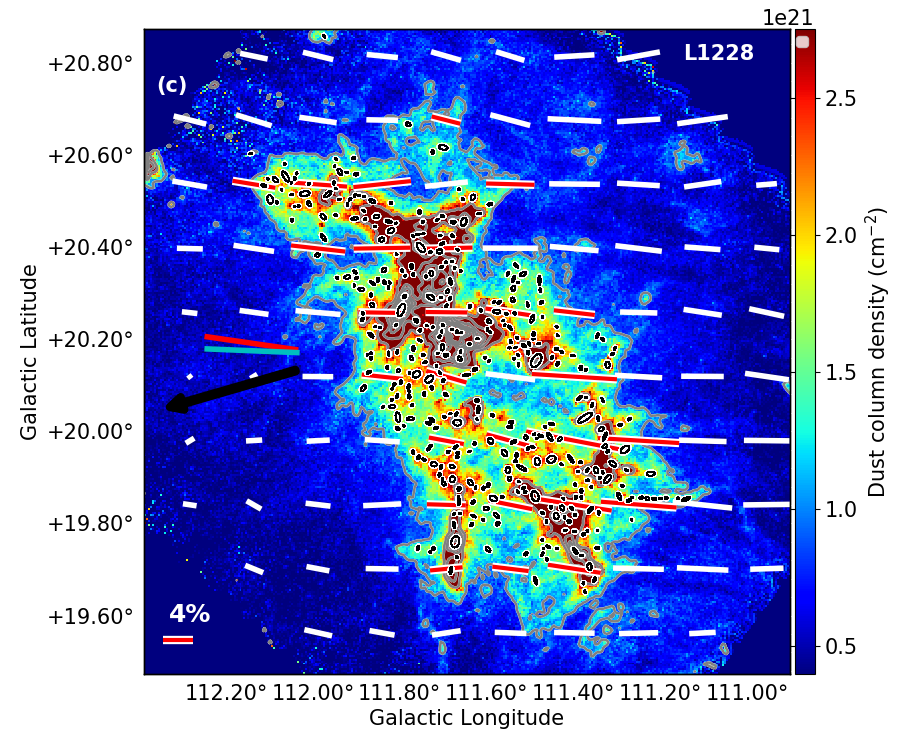}
   \includegraphics[width=8.9cm, height=6cm]{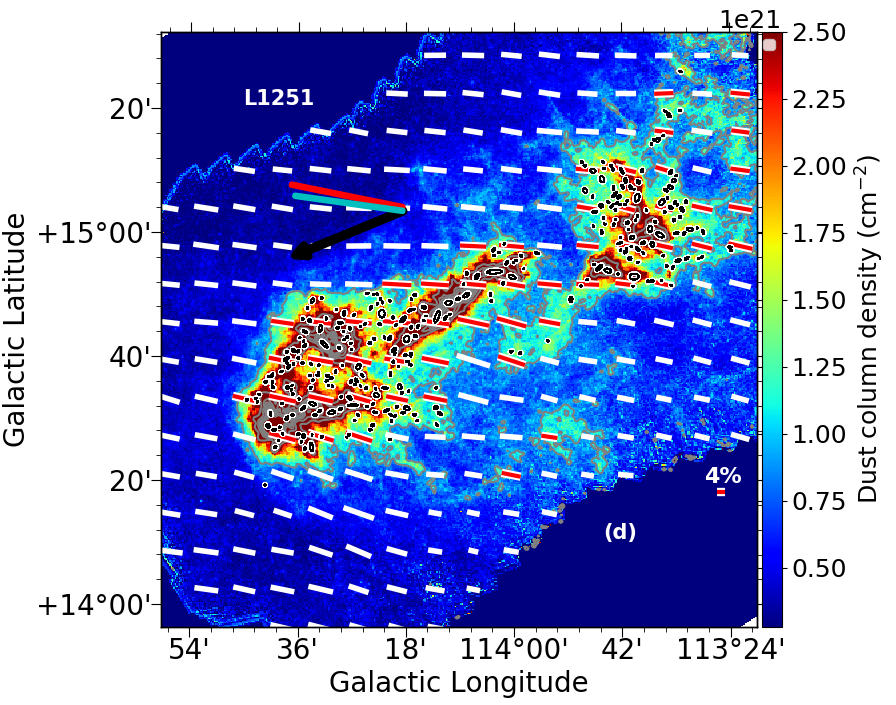}
  \caption{Planck polarization vectors (white) plotted over the column density maps of the clouds produced using the \textit{Herschel} images. The vectors identified with the red+white vectors corresponds to a column density of $1\times10^{21}$ cm$^{-2}$ which are considered as representing the cloud magnetic field geometry. The arrow drawn in black shows the direction of motion of the clouds. The thick lines drawn in red and cyan are the mean value of the cloud and the background magnetic field orientations respectively.}\label{fig:pol-plk-gal}
\end{figure*}
The plane of the sky magnetic field (B$_{pos}$) vectors obtained from the \textit{Planck} polarization measurements of the region containing all the four complexes are shown in Fig. \ref{fig:pol-plk-gal}. The B$_{pos}$ vectors are overplotted on the \textit{Herschel} column density maps. The outer most contour drawn in Fig. \ref{fig:pol-plk-gal} corresponds to a column density of $1\times10^{21}$ cm$^{-2}$. This corresponds to an extinction of A$_{V}\approx0.5$ magnitude determined using the relationship between the column density and the extinction derived by \citet{2003ARA&A..41..241D} for R$_{V}=$3.1 extinction law.  While the vectors lying within the contour are considered as providing information on the cloud B$_{pos}$, the ones lying outside are considered as providing information on the ambient or inter-cloud magnetic field (ICMF) orientation surrounding each cloud. The B$_{pos}$ vectors lying within (B$^{cloud}_{pos}$) and outside (B$^{ICMF}_{pos}$) of this contour are identified separately using the lines drawn in red+white and white colors respectively. 

We found no significant difference between the distributions of B$^{cloud}_{pos}$ and B$^{ICMF}_{pos}$ vectors in L1147/1158, L1228 and L1251 which is very much evident from Fig. \ref{fig:pol-plk-gal} also. This implies that the clouds are threaded by the ambient magnetic fields. We found a change in the orientation of the B$^{cloud}_{pos}$ and B$^{ICMF}_{pos}$ vectors of L1172/1174 which is more prominent as we move to the south and the south-eastern direction of the cloud. The orientation of the B$^{cloud}_{pos}$ and B$^{ICMF}_{pos}$ for the clouds are obtained by calculating the median value of all the vectors lying within and outside of the $1\times10^{21}$ cm$^{-2}$ contour. The values of $\theta^{cloud}_{Bpos}$ and $\theta^{ICMF}_{Bpos}$ thus obtained are listed in column 10 of Table \ref{tab:yso_dist_pm_mean}. The $\theta^{cloud}_{Bpos}$ and the $\theta^{ICMF}_{Bpos}$ determined for the clouds are shown using thick lines in red and cyan in Fig. \ref{fig:pol-plk-gal}. 


\subsubsection{L1147/1158}
In L1147/1158, we noted that the cloud complex has strikingly sharp edges to its southern side which is the side facing the Galactic plane. The sharp edges of the southern side of L1147/1158 were also noted by \citet{1991A&A...249..493H}. The projected direction of motion of L1147/1158 is found to be $\sim$157\degr~with respect to the Galactic north. The results imply that L1147/L1158 is moving towards the Galactic plane. Interestingly, the sharp edges are developed towards the leading edge of the cloud motion. It is possible that these sharp edges are created as a result of the interaction of the cloud with the ambient medium through which it is travelling. The projected magnetic field orientations found inside and outside the cloud boundary are found to be at an angle of 183$\pm$7\degr~and 186$\pm$6\degr, respectively. The directional offset between the inner and outer magnetic fields is found to be 3\degr. The projected offsets between the direction of motion of the cloud and the magnetic field orientation inside and outside the cloud boundary are 26\degr~and 29\degr~respectively. Thus the projected motion of the cloud is almost aligned with the ambient magnetic field direction.

\subsubsection{L1172/1174}
L1172/1174 which seems to be both spatially and kinematically associated with L1147/1158 shows a significant difference in the projected magnetic field orientation especially to its northern and southern regions. The field orientation to the north of L1172/1174 is consistent with the field orientation seen towards L1147/1158. However, it is to the south-eastern side of L1172/1174 that the magnetic field orientation becomes more complex. The projected magnetic field orientations inside and outside of the cloud are 208$\pm$11\degr~and 196$\pm$23\degr~respectively. The directional offset between the inner and outer magnetic fields in L1172/1174 is found to be 12\degr. The complex magnetic field orientation in the vicinity of the complex is evident from the relatively high dispersion (by a factor of two to three compared to the values found towards the other three clouds) found for the field vectors lying outside the cloud. Similar to L1147/L1158, L1172/1174 is also found to be moving towards the Galactic plane. The projected direction of motion of L1172/1174 is estimated to be 151\degr~which is quite consistent with the direction of the motion of L1147/1158. Thus, the offsets between the direction of motion of the cloud and the magnetic field orientations found inside and outside of the cloud are 57\degr~and 45\degr~respectively. 

\subsubsection{L1228}
Among the four clouds studied here, L1228 is located at the farthest distance of $\sim$121 pc from the Galactic mid-plane. The direction of motion of the L1228 is found to be at an angle of 107\degr~with respect to the Galactic north. The projected orientation of the magnetic field inside and outside of L1228 are found to be 84$\pm$7\degr~and 88$\pm$10\degr~respectively. The offset angle between the inner and outer magnetic fields in L1228 is found to be 4\degr. The projected direction of motion is at an offset of 23\degr~and 19\degr~with respect to the projected magnetic fields within and outside of the cloud respectively. 

\subsubsection{L1251}
L1251 is an elongated cloud having a remarkable cometary morphology which is believed to have been formed as a consequence of its interaction with a supernova bubble as described by \citet{1989ApJ...347..231G}. The elongation of the cloud is in the east-west direction. Two IRAS point sources, associated with molecular outflows, are located in the eastern head region of the cloud while no sign of current star formation was identified in the western region which forms the tail of the cloud \citep[e.g., ][]{2008hsf1.book..136K}. Based on the morphology of the cloud produced using the extinction maps generated from the star counts, \citet{2004A&A...425..133B} suggested that the cloud resembles a body flying at hypersonic speed across an ambient medium. The projected direction of motion of the cloud is found to be at a direction of 107\degr. The projected orientation of the magnetic field inside and outside of L1251 is found to be 79$\pm6$\degr~and 81$\pm6$\degr~respectively and the angular offset between the two fields is 2\degr. The offset between the projected directions of the motion of the cloud and the inner and the outer magnetic fields are 28\degr~ and 26\degr~. It is interesting to note that while the cometary morphology of the cloud is oriented parallel to the direction of motion.

The inner and outer magnetic fields in all the four clouds are found to be parallel to each other suggesting that the inner magnetic fields are inherited from the ambient fields and that the formation of the clouds have not affected the cloud geometries much. In a magnetic field mediated cloud formation scenario, the field lines guide the material \citep{1999ApJ...527..285B, 2014ApJ...789...37V,2010ApJ...723..425D}, to form the filamentary structures that are expected to be oriented perpendicular to the ambient magnetic fields which then undergo fragmentation to form cores \citep{2013ApJ...777L..33P, 2015A&A...584A..91K,2020A&A...642A.177D}. As the accumulation of the material is helped by the magnetic field lines, the ICMF direction is expected to get preserved deep inside the cores \citep{2009ApJ...704..891L, 2013ApJ...768..159H, 2015Natur.520..518L}, expecting a parallel orientation between the ICMF and the core magnetic field.

The magnetic field lines aligned with the cloud motion was studied in 2D numerical simulations by \citep{1994ApJ...433..757M}. They showed that the magnetic fields help the cloud to stabilize against disruptive instabilities. Further, using 2D numerical study, \citet{1999ApJ...517..242M} explored the effect of a uniform magnetic field oriented oblique with respect to a moving interstellar cloud. They carried out the study for several values of inclination ranging from the aligned case to the transverse case, for several values of Mach number and density contrast parameter which is the ratio between the cloud and the ambient density. They found that for angles larger than 30\degr, the magnetic field lines get stretched and would tend to wrap around the cloud amplifying the magnetic field significantly at the expense of the kinetic energy of the cloud.

In all the clouds, the field orientations are smooth and well ordered. In L1148/1157, L1228 and in L1251, the offsets between the projected direction of the motion and the projected magnetic fields are $\lesssim$30\degr. No draping of the magnetic fields towards the leading edges of the clouds are noticed even in the case of the cometary cloud L1251. However, the offset between the projected direction of the motion and the projected magnetic fields in L1172/1174 is $\gtrsim$30\degr~and one would expect an amplification of the field strength towards its leading edge. Among the four clouds studied here, L1172/1174 is the only cloud that shows an apparent hub-filament structure and forming a massive star, HD200775, and a sparse cluster in the hub. It is interesting to note that in L1172/1174, the deviation in magnetic field as we move from L1147/1158 occurs at the location of the cloud and then as we move towards the eastern/south-eastern sides of L1172/1174, the field geometry becomes complex. It could be possible that a large offset between the cloud motion and the magnetic fields might have helped the cloud to amass material quickly and initiate the formation of a massive star like HD200775. Whether the modification of the magnetic fields occur due to the formation of the cloud or the bend in the magnetic field facilitated the formation of the cloud is unclear.

\subsection{Magnetic field strength}
It is also useful to estimate the magnetic field strength in the four clouds. We used Davis-Chandrasekhar \& Fermi (DCF) \citep{1951PhRv...81..890D,1953ApJ...118..113C} method for the calculation of the plane-of-sky magnetic field strength. The DCF method is given as,
\begin{equation}
B_{pos}= 9.3 \times \sqrt{\frac{n_{H_{2}}}{cm^{-3}}} \times  \frac{\rm \Delta v}{km s^{-1}} \times  \Big(\frac{\Delta \phi}{1^{\circ}}\Big)^{-1},
\end{equation}
where $\rm\Delta$v is the velocity dispersion or Full Width Half Maxima (FWHM) of spectral line in the units of \kms, n$_{H_{2}}$ is the volume density and $\Delta\phi$ is the dispersion in polarization angles which is calculated using Stokes parameters as follows:
\begin{align}
\Delta \phi = \sqrt{\langle(\Delta \psi)^{2}\rangle},
\end{align}
\begin{align}
\Delta \psi = \frac{1}{2} \arctan(Q\langle U\rangle - \langle Q\rangle U,Q\langle Q\rangle+\langle U\rangle U), 
\end{align}
where $\langle$Q$\rangle$ and $\langle$U$\rangle$ are the average of stokes parameters over the selected pixels \citep{2015A&A...576A.104P}.

We obtain the velocity dispersion from the most complete CO survey done using NANTEN telescope with spectrum at every 1/8\dgr part of the molecular cloud with the velocity resolution of 0.24 \kms. Considering the large difference between beam size of CO and \textit{Planck} data, we selected only those positions in the cloud where $\abs{U}$/$\sigma_{U}>3$ and $\abs{Q}$/$\sigma_{Q}>3$. This analysis has been adopted from \cite{2016A&A...586A.138P}. The number density used in the calculation requires assumption of a particular cloud geometry for all the four clouds which will result in the additional uncertainties. Here, for simplicity, we assume a number density of 100 cm$^{-3}$ for all the clouds, which is a typical value of molecular clouds studied here \citep{2011piim.book.....D}. 
The dispersion in the polarization angles ($\Delta\phi$) using the stokes parameters at the chosen pixels is 5\dgr-13\dgr and the mean velocity width (full width at half maxima) using $^{12}$CO observations is 2.1-2.7 \kms. Only those spectra were chosen that are showing a single Gaussian at the selected positions from Stokes parameter maps. Since there is higher dispersion in polarization angles in the case of L1172/1174 over the head part due to the presence of a star HD 200775, we took into account the polarization angles over the tail part of the cloud where the dispersion is due to small scale variations. The values of the magnetic field strength calculated in the four clouds are 23, 18, 26 and 44 $\mu$G for L1147/1158, L1172/1174, L1228 and L1251 respectively. The typical uncertainties in the estimation of magnetic field strength are $\sim$ 0.5 B$_{pos}$ as found in earlier studies \citep{2004Ap&SS.292..225C,2016A&A...586A.138P}. This implies that the calculated values of magnetic field strength in all the clouds are comparable within the uncertainty.



\subsection{Relative orientations of clumps, magnetic field and motion of the clouds}

\begin{figure}
   \includegraphics[width=4.3cm, height=5cm]{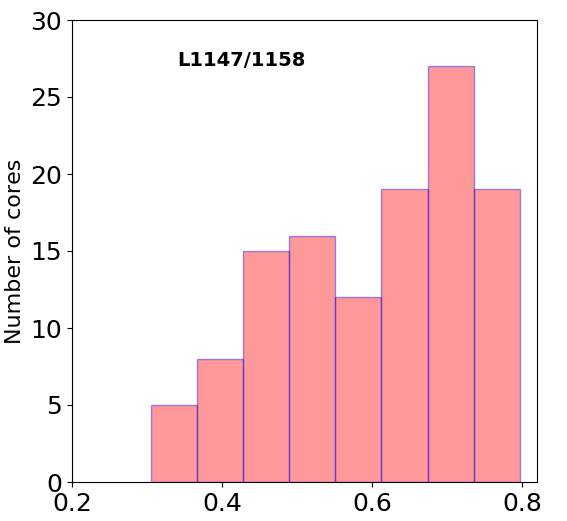}
   \includegraphics[width=4.3cm, height=5cm]{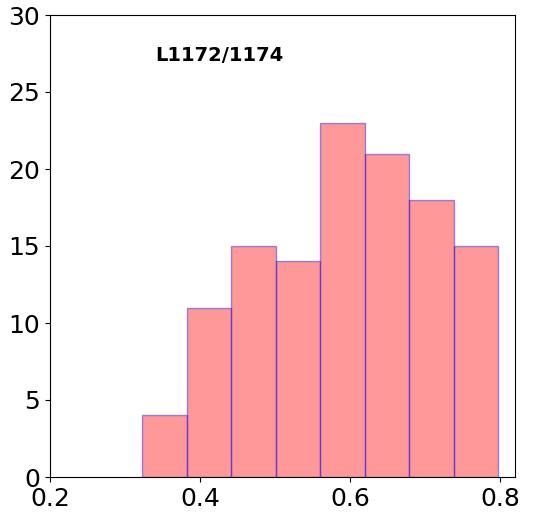}
   \includegraphics[width=4.4cm, height=5cm]{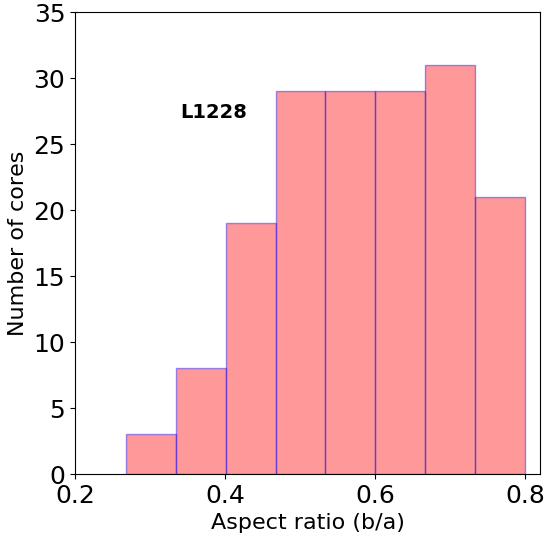}
   \includegraphics[width=4.3cm, height=5cm]{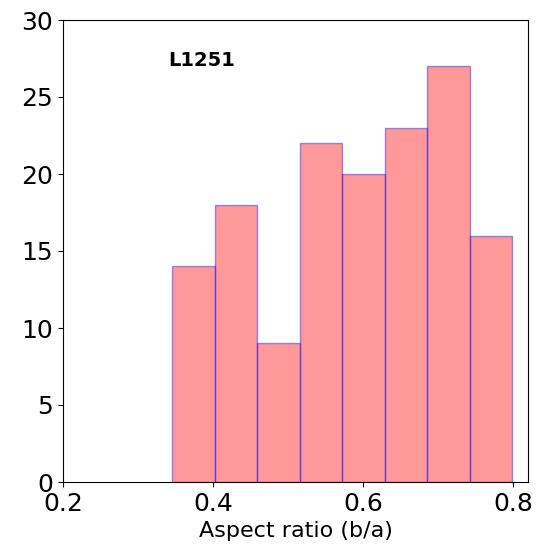}
  \caption{Histogram of the aspect ratios of the clumps extracted from the four clouds.}\label{fig:hist_aspect}
\end{figure}

The clumps extracted by \cite{2020ApJ...904..172D} in all the four clouds are mostly associated with the filamentary structures which is consistent with the results obtained  from the Herschel Gould Belt Survey \citep{2010A&A...518L.102A, 2010A&A...518L.106K, 2014prpl.conf...27A}. Theoretically, this association is interpreted as the longitudinal fragmentation of thermally supercritical filaments into cores \citep[e.g., ][]{1992ApJ...388..392I, 1997ApJ...480..681I}. However, in the presence of turbulence, the filaments cease to become quiescent structures in which perturbations grow slowly. In the case when the clouds are in motion through the ambient ISM, like the four clouds studied here, they interact with it generating turbulence in the cloud structure \citep{1994ApJ...433..757M, 1996ApJ...473..365J, 1999ApJ...517..242M, 1999ApJ...527L.113G, 2000ApJ...543..775G}. Under this scenario we made an attempt to examine the properties of the clumps associated with these clouds with respect to the directions of the $\theta^{cloud}_{Bpos}$ and $\theta^{ICMF}_{Bpos}$ and the projected direction of their motion, $\theta^{motion}_{pos}$. The distribution of the aspect ratio of the identified clumps in each cloud are shown in Fig. \ref{fig:hist_aspect}. The clumps in all the clouds show whole range of aspect ratio from 0.3 to 0.8. A majority of cores in all panels  ($\sim$ 75$\%$) show aspect ratio greater than 0.5 which implies that a higher number of sources exhibits a less flattened geometry. Since this study is based on projected maps, the less flattening of cores could also be a consequence of projection effects from 3D to 2D plane \citep{2020MNRAS.494.1971C}.

On the basis of the ratio of the mass of Bonner-Ebert (BE) sphere \citep{1956MNRAS.116..351B,1955ZA.....37..217E} and the mass of core (M$_{core}$/M$_{BE}$), the whole sample for four clouds was classified into prestellar and starless cores by \cite{2020ApJ...904..172D}. The cores having ratio less than 2 are considered as starless cores whereas the ones with ratio higher than 2 are considered as gravitationally bound termed as candidate-prestellar cores and may collapse to form stars. 
Assuming the different evolution of the cores in their prestellar and starless stage, we showed their respective distribution separately. The starless cores, being gravitationally unbound are at earlier stage of evolution tend to be more elongated and less spherical shape as compared to the prestellar core evolution. We filtered the sources for the aspect ratio less than 0.8 to consider only elongated/asymmetric sources. Hence, the final sample of prestellar and starless sources for L1147/1158 is 46 and 75; for L1172/1174 it is 38 and 83; for L1228, it is 54 and 115; for L1251, it is 69 and 89. We notice that more number of prestellar cores gets filtered out as compared to starless cores which implies that the prestellar cores are more spherical as compared to starless sample in all the four clouds.
\begin{figure*}
   \includegraphics[width=9.1cm, height=6.0cm]{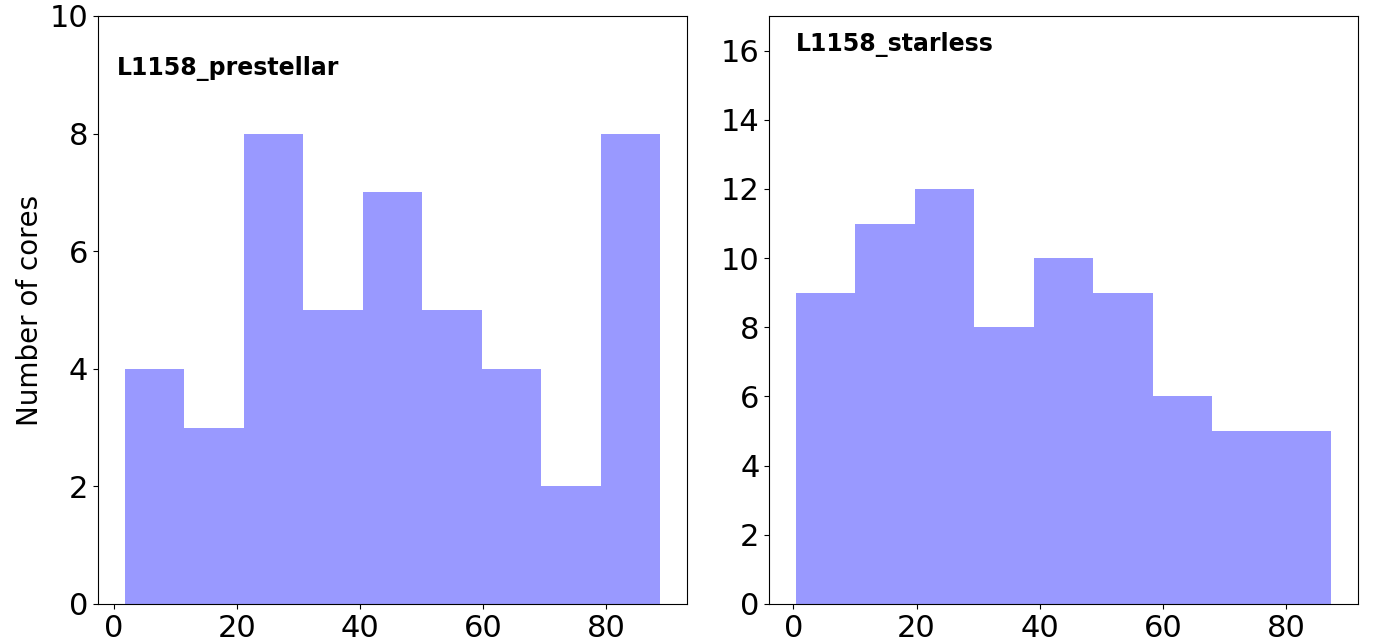}
   \includegraphics[width=9.1cm, height=6.0cm]{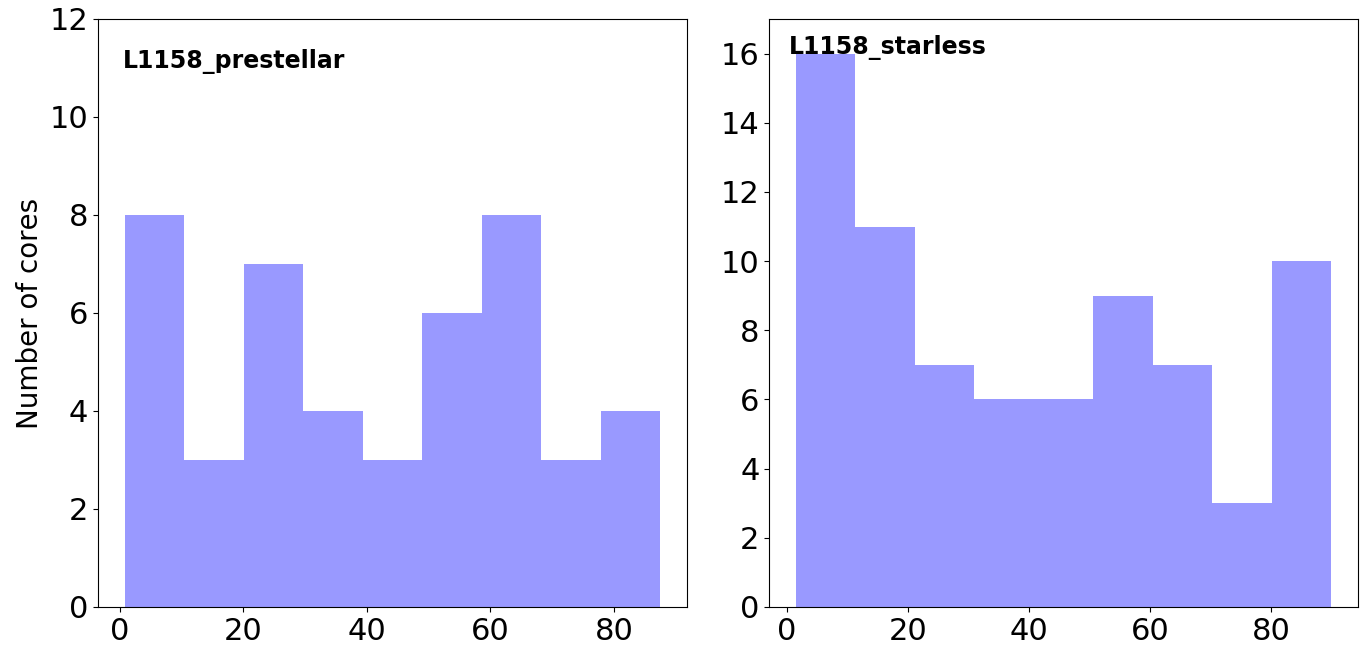}
   \includegraphics[width=9.1cm, height=6.0cm]{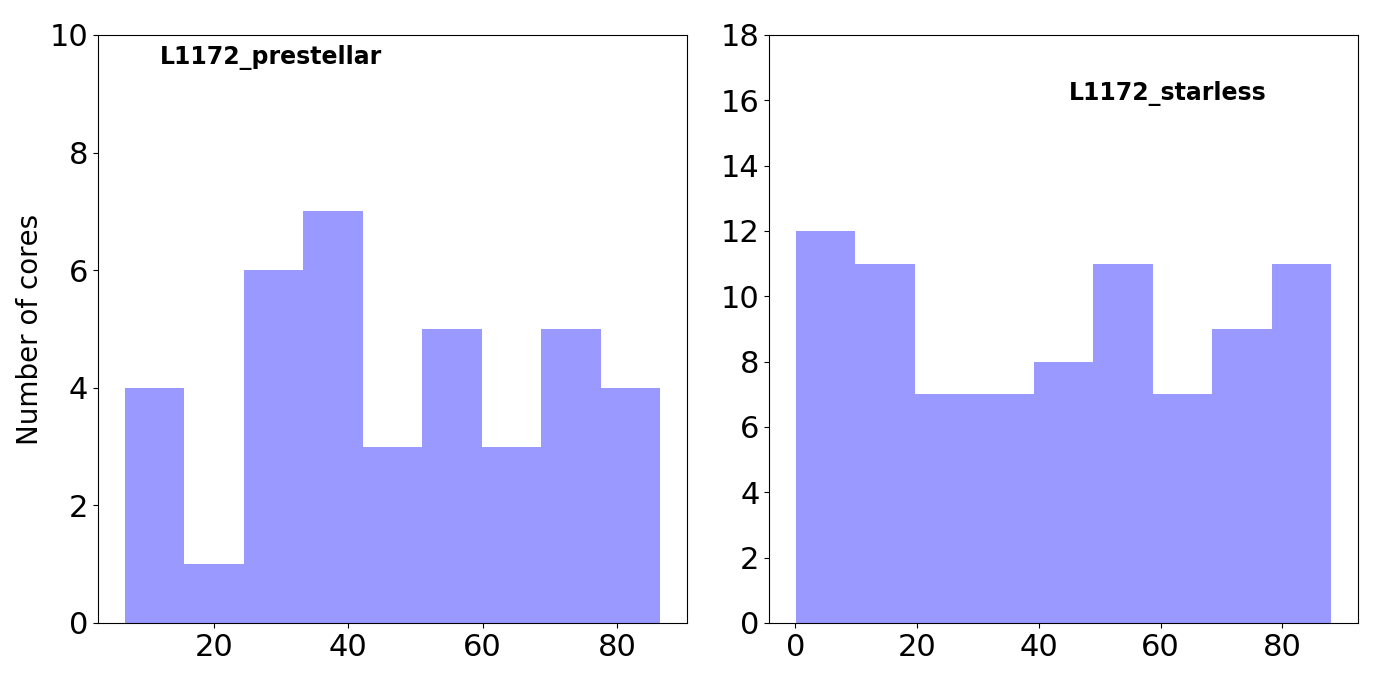}
   \includegraphics[width=9.1cm, height=6.0cm]{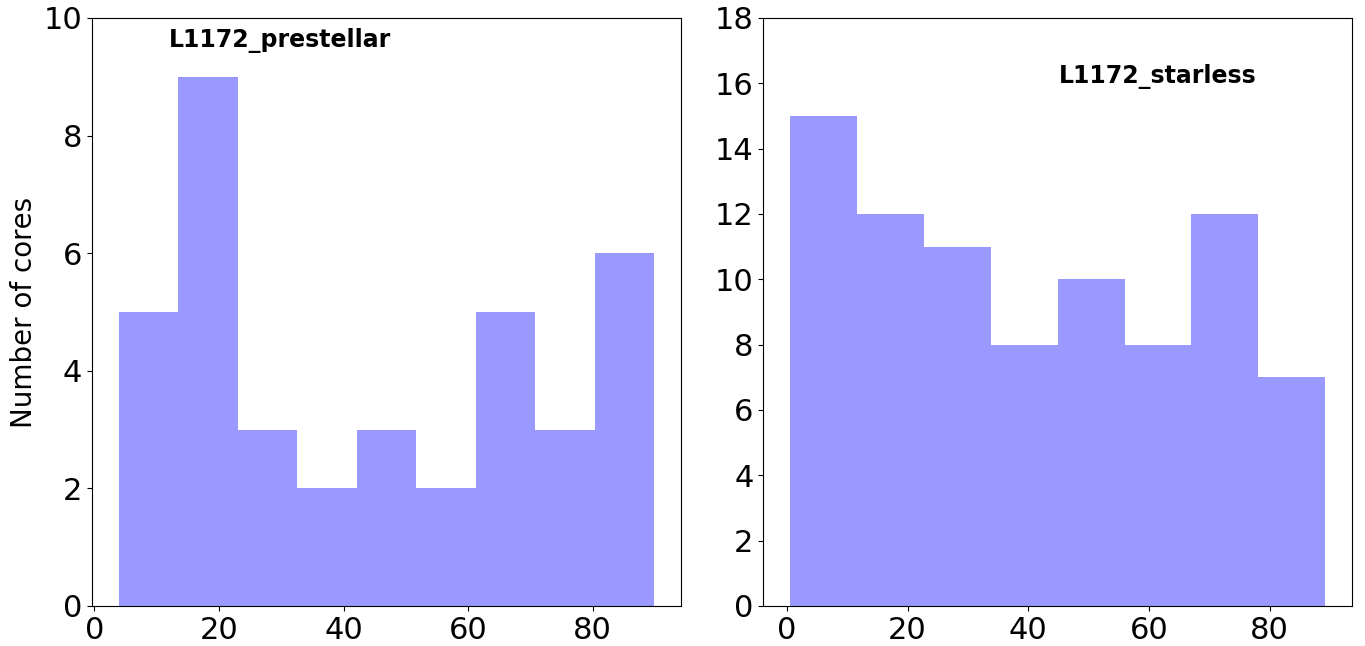}
   \includegraphics[width=9.1cm, height=6.0cm]{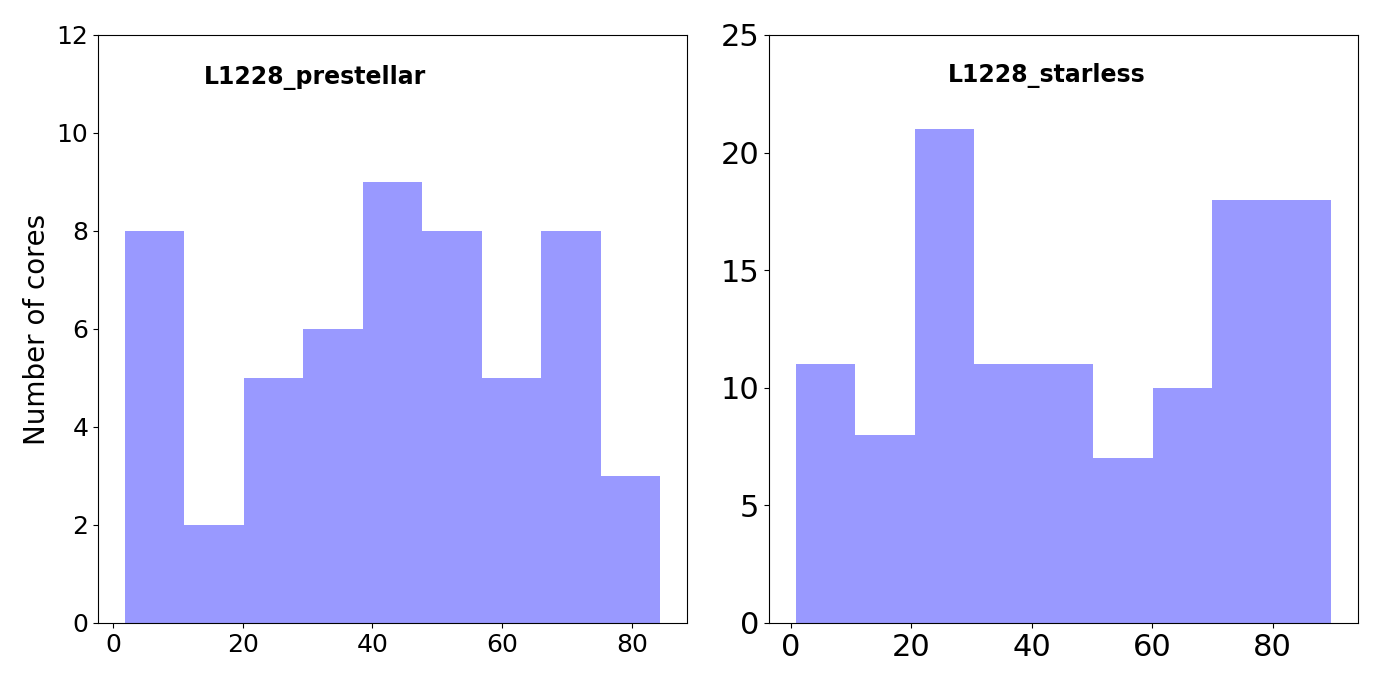}
   \includegraphics[width=9.1cm, height=6.0cm]{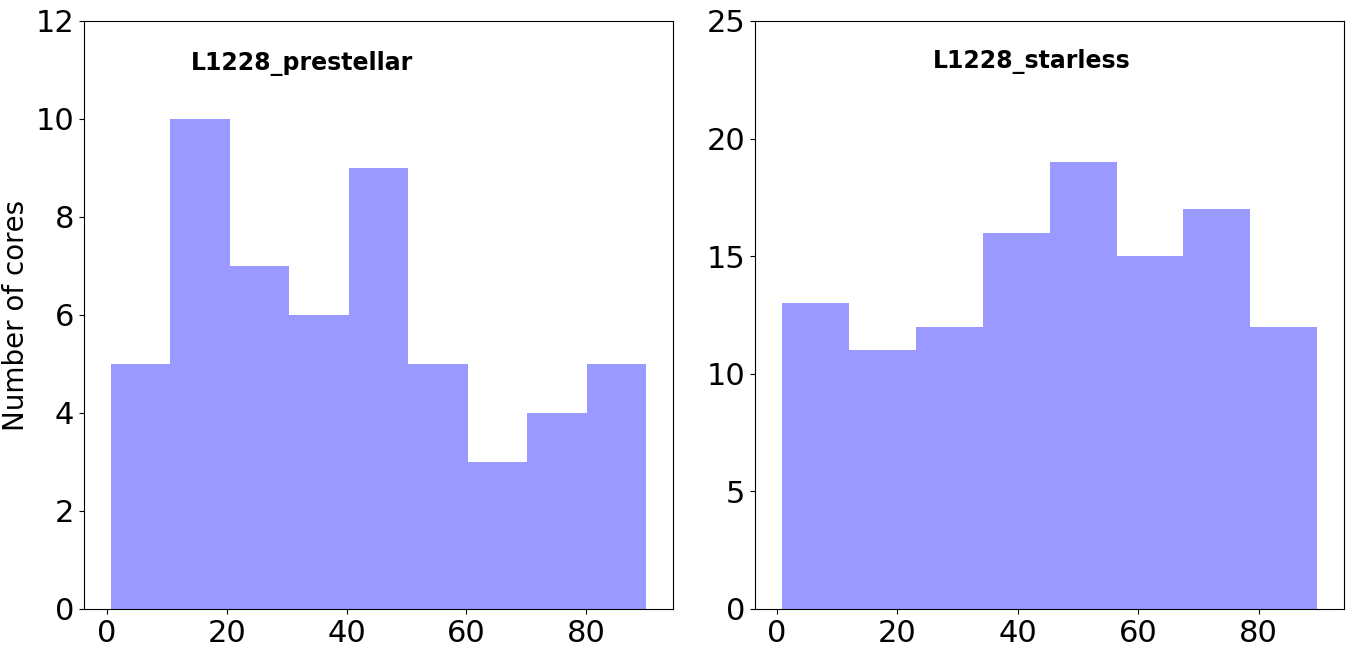}
   \includegraphics[width=9.1cm, height=6.0cm]{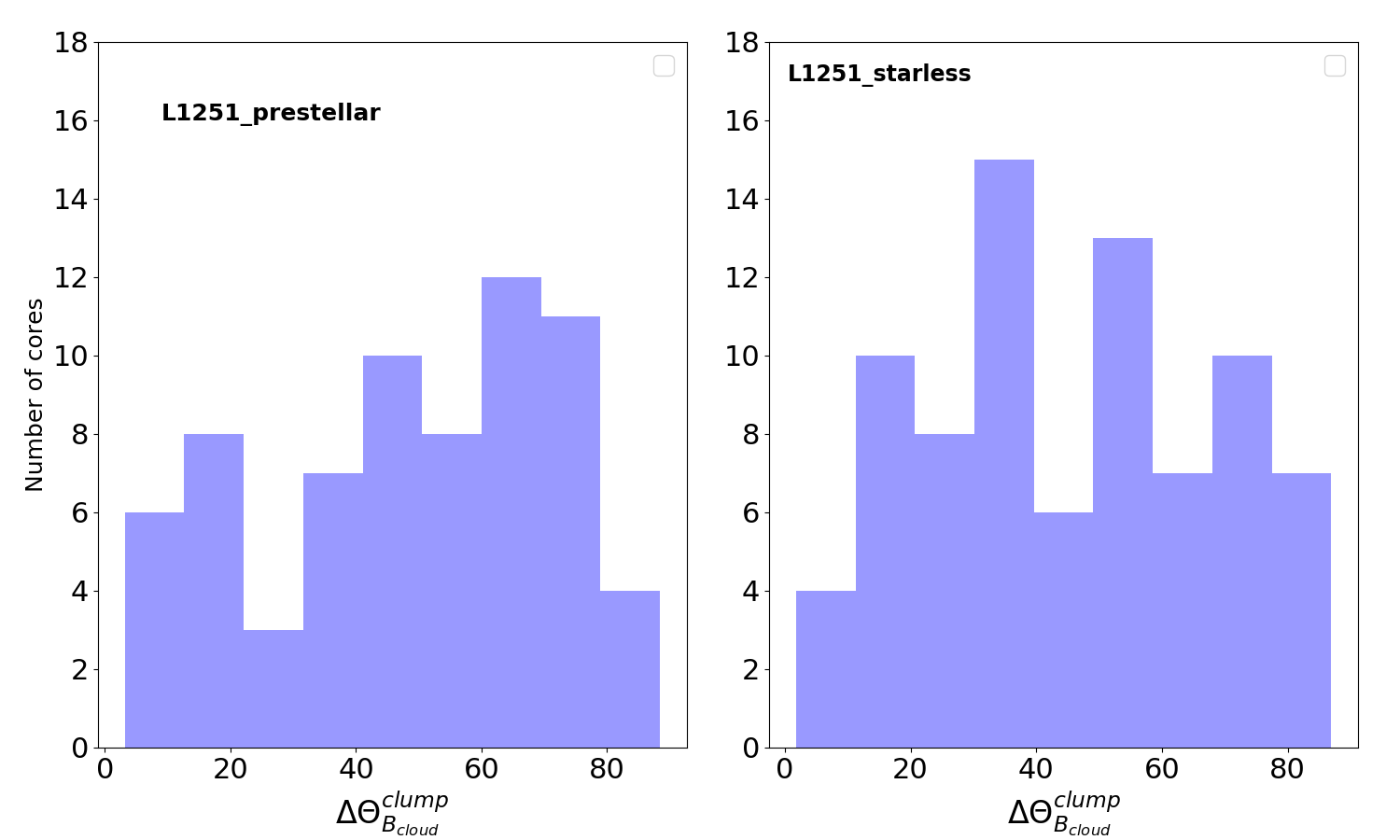}
   \includegraphics[width=9.1cm, height=6.0cm]{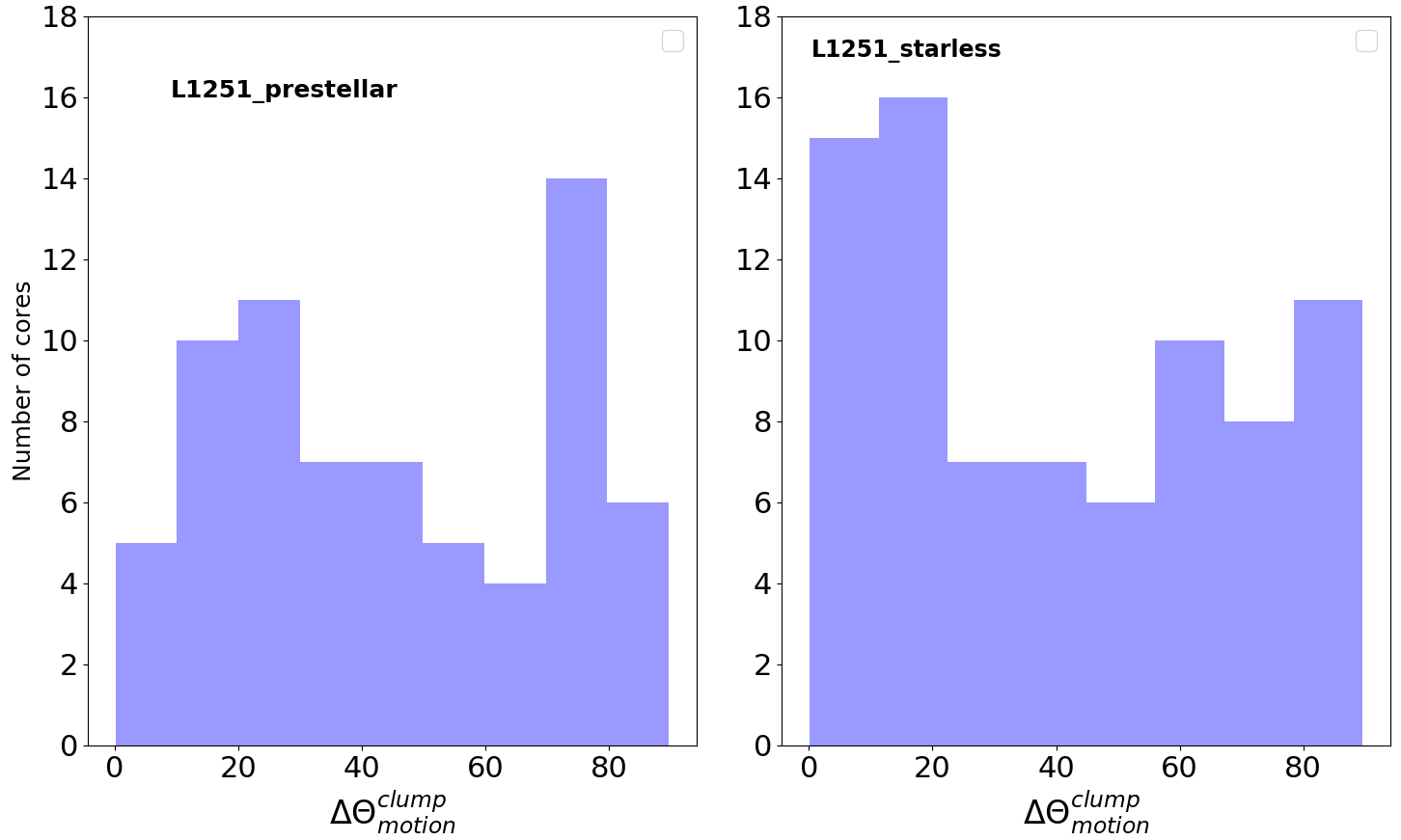}
  \caption{{\bf Left (two columns):} Histograms of the offsets for all the four clouds between $\theta^{clump}_{pos}$ with respect to the $\theta^{cloud}_{Bpos}$ (blue).  
  {\bf Right (two columns):} $\theta^{clump}_{pos}$ with respect to $\theta^{motion}_{pos}$. }\label{fig:hist_clumag}
\end{figure*}


  We used the position angles of major axis for all the cores extracted from high resolution column density map at 18.2$\arcsec$ resolution \citep{2020ApJ...904..172D}. The offset angles, $\Delta\theta^{clump}_{Bcloud}=$ $\theta^{clump}_{pos}$ - $\theta^{cloud}_{Bpos}$, $\Delta\theta^{clump}_{BICMF}=$ $\theta^{clump}_{pos}$ - $\theta^{ICMF}_{Bpos}$, and $\Delta\theta^{clump}_{motion}=$ $\theta^{clump}_{pos}$ - $\theta^{motion}_{pos}$ are calculated. We show here only $\Delta\theta^{clump}_{Bcloud}$ since the values of $\Delta\theta^{clump}_{BICMF}$ and $\Delta\theta^{clump}_{Bcloud}$ are comparable within the uncertainties. This shows that the direction of the magnetic field is preserved from cloud to inter-cloud scales. The distributions of $\Delta\theta^{clump}_{Bcloud}$ for all the four clouds are shown in Fig. \ref{fig:hist_clumag} on the left side whereas $\Delta\theta^{clump}_{motion}$ is shown on the right side.
Towards L1147/1158, for the prestellar cores, the major axis of the clumps are oriented in the range of 30\degr-60\degr with respect to the ICMF and the inner magnetic fields. In case of starless cores, both $\Delta\theta^{clump}_{BICMF}$ and $\Delta\theta^{clump}_{Bcloud}$ are found to lie mostly $\leq30$\degr. The starless cores show a preferred parallel orientation with respect to magnetic field.  In L1172/1174, the major axis of the clumps are random to the cloud magnetic field and the ICMF. Out of 38 prestellar cores, the distribution is 11, 14 and 13 clumps $\Delta\theta^{clump}_{Bcloud}<$30\degr, 30\degr~$<\Delta\theta^{clump}_{Bcloud}\leqslant$60\degr~and $\Delta\theta^{clump}_{motion}$ $>60$\degr ~respectively. Of the 83 starless clumps identified in L1172/1174, 30, 29 and 24 clumps are present in three ranges implying a uniform distribution. The pattern of $\Delta\theta^{clump}_{BICMF}$ is also found to be similar to that of $\Delta\theta^{clump}_{Bcloud}$. In L1228, the offsets, $\Delta\theta^{clump}_{BICMF}$ and $\Delta\theta^{clump}_{Bcloud}$ show a preferred direction around 40\degr-60\degr with higher number of 20 (37\%) sources in this specific range. On the contrary, the distribution is random for the starless cores showing a lack of any preferred orientation. For L1251, there is again a random distribution for starless as well as prestellar cores. 

In L1147/1158, the offsets for prestellar cores between the clump major axis and the motion of the cloud, $\Delta\theta^{clump}_{motion}$ is random whereas in starless cores, more number of cores are lying $\leq$30\degr implying that more clumps are parallel to the projected direction of motion. 
In L1172/1174, out of 38 prestellar clumps, the offsets show bimodal distribution with 17 clumps lying at $<$30\degr and 14 at $>$60\degr. The offsets in starless cores lie mostly at less than 30\degr which implies that the clumps are predominantly parallel to the direction of motion. In L1228, the distribution of $\Delta\theta^{clump}_{motion}$ for prestellar cores seems to be more random while for starless cores, there is a preferred orientation with more number of clumps lie within range of 30\degr-60\degr. In L1251, out of 69 prestellar clumps, 26 (38\%) show the values of $\Delta\theta^{clump}_{motion}\leq$30\degr~implying that the major axis of the clumps are aligned with the direction of the motion of the cloud. In fact, the cloud also shows a blunt head and a tail which is filamentary and is aligned almost parallel to the direction of the motion. It is also important to notice here that the major axis of the majority of the clumps is also aligned along both the filament and the direction of the motion of the cloud. Out of 80 starless clumps, the offset distribution shows a clear bimodal trend where 36 (45\%) sources lie below 30\degr and 27 (33\%) sources greater than 60\degr. 


\begin{figure*}
     \includegraphics[width=9cm, height=6cm]{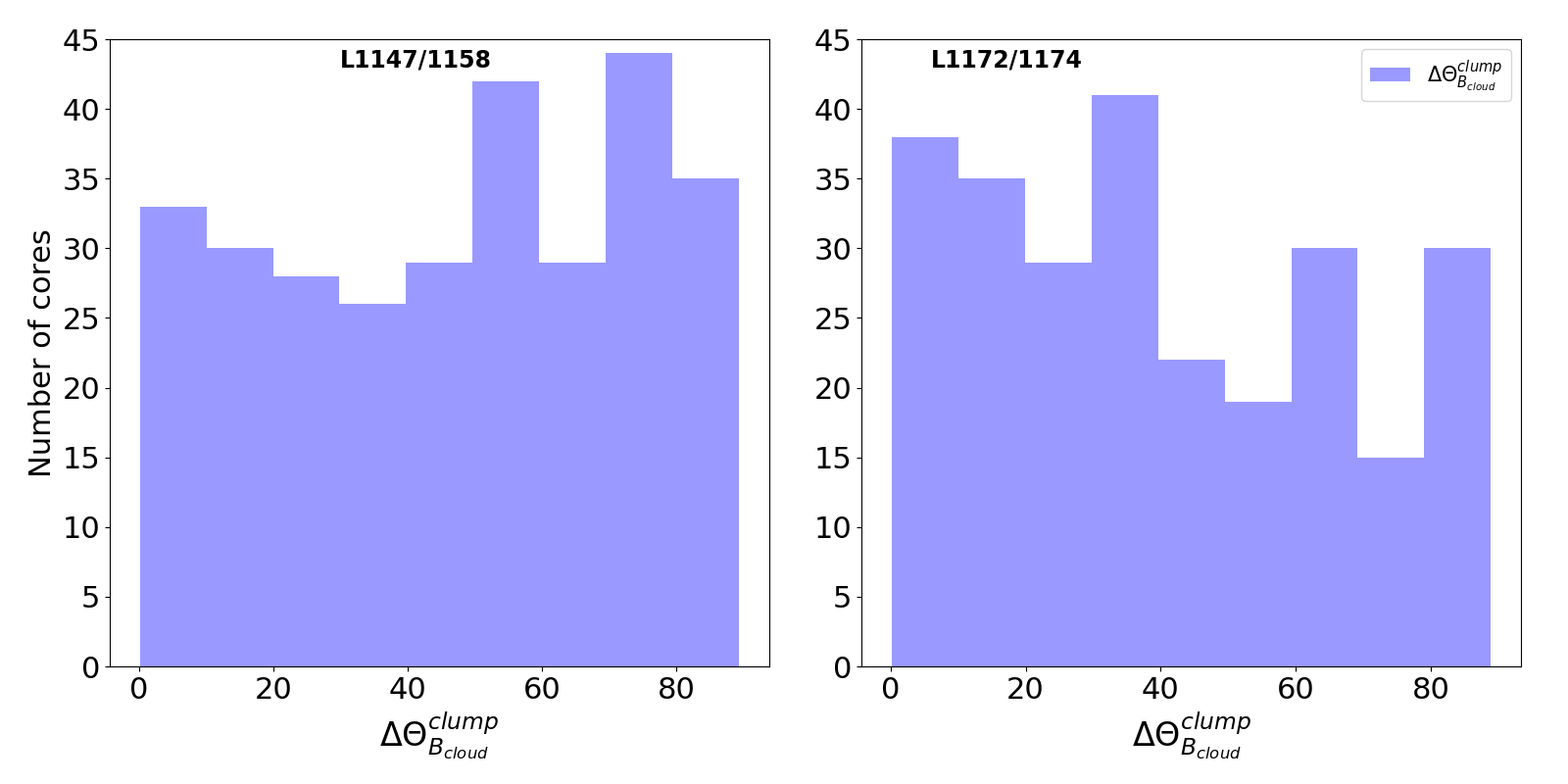}
     \includegraphics[width=9cm, height=6cm]{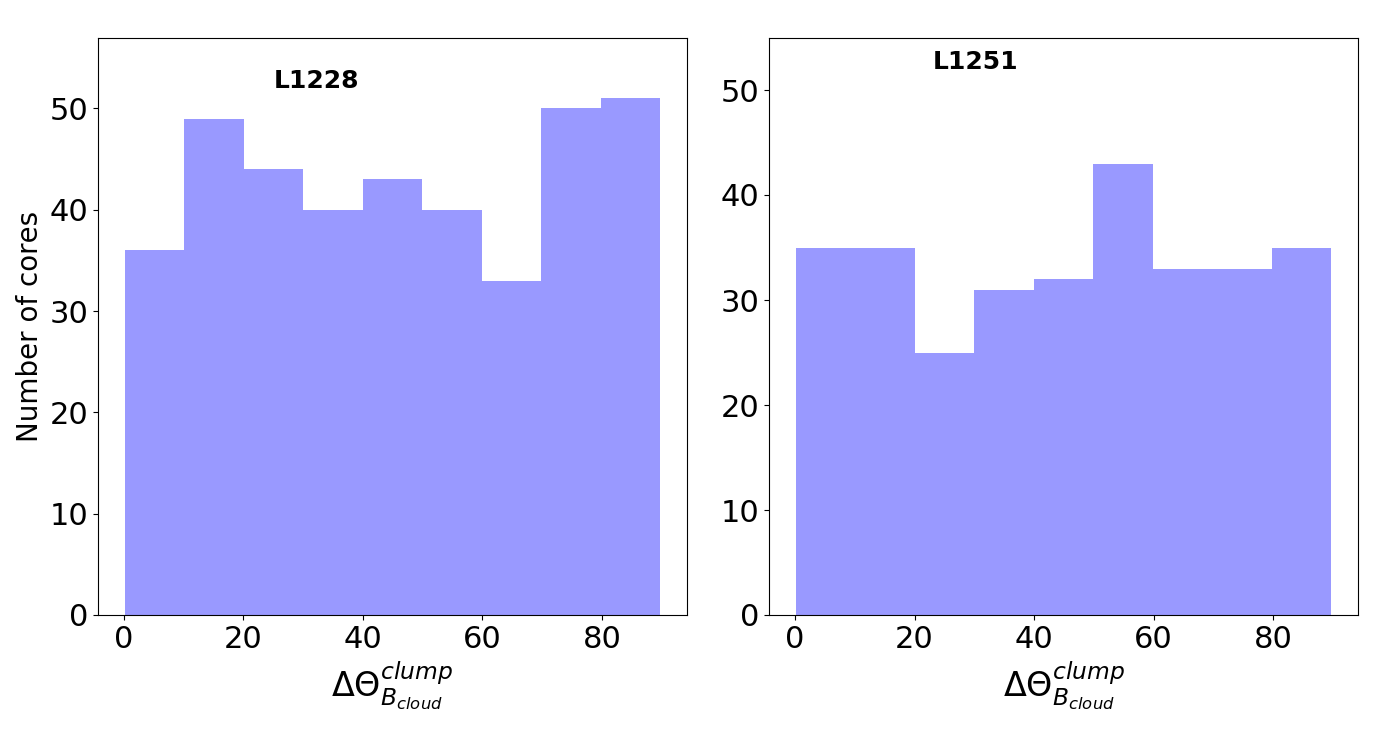}
     \vspace{0.2cm}
     \includegraphics[width=9.2cm, height=6cm]{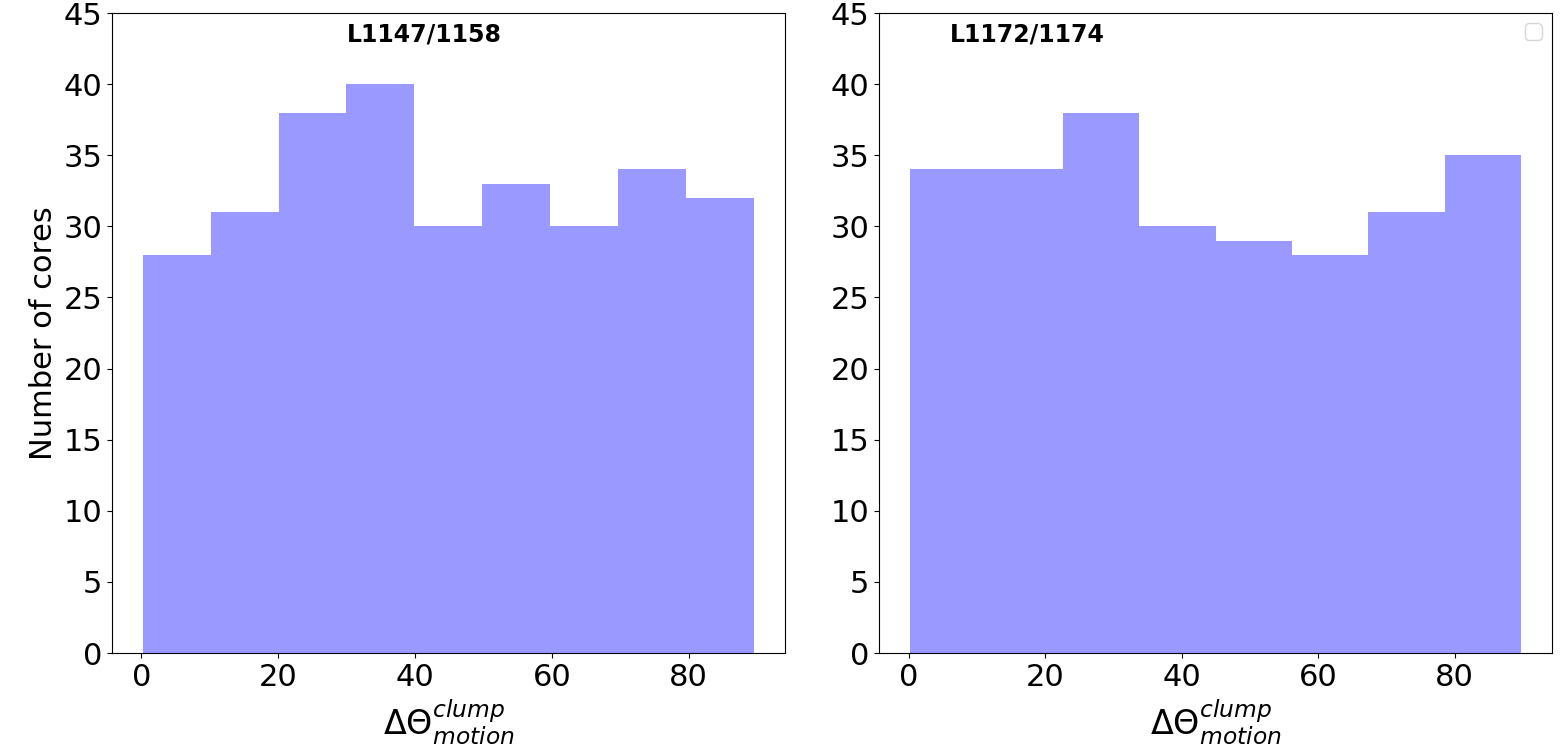}
     \includegraphics[width=9cm, height=6cm]{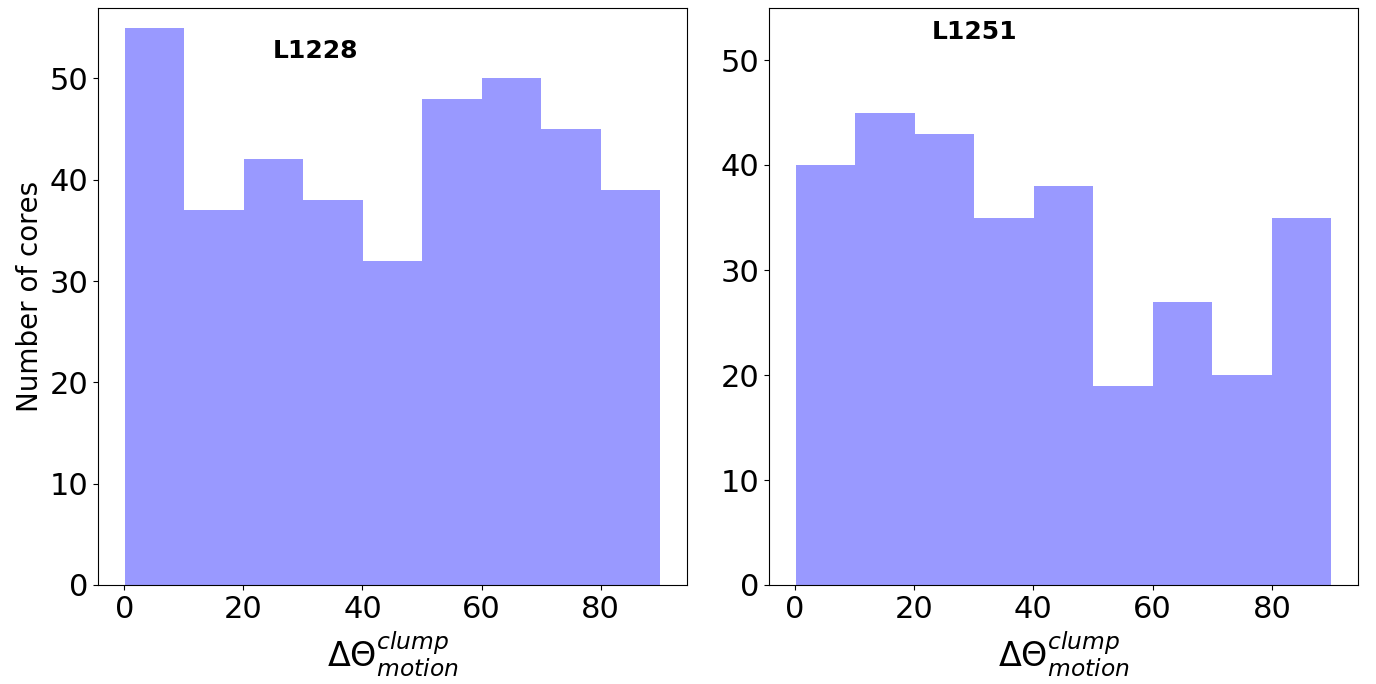}
  \caption{{\bf Upper row:} Histogram of the offsets between inner/outer magnetic field with the clump orientation. {\bf Bottom row:} Histogram of the offsets between the direction of motion with the clump orientation.}\label{fig:hist_clumot}
\end{figure*}



As mentioned in section \ref{ide_clump}, to check any dependency on the nature of source extraction method, we have used \textit{Astrodendro} to extract the clumps in all the four clouds. The parameters used for the extraction have been discussed in section \ref{ide_clump}. We derived their properties like the right ascension and the declination of the identified clumps, their effective radius, major and minor axis and the position angle of the major axis of the clumps ($\theta^{clump}_{pos}$). The reference is taken from galactic north counter-clockwise although the position angle in the \textit{Astrodendro} is given from the positive x-axis increasing counter-clockwise. We extracted 352 clumps each in L1147/1158 and 295 in L1172/1174, 437 in L1228 and 345 in L1251 complex. After applying all the selection criteria on the size and the aspect ratio mentioned in the section \ref{ide_clump}, the final number of clumps are 296 in L1147/1158, 259 in L1172/1174, 386 in L1228 and 302 in L1251 cloud complex.  While deriving the sources, we ignored the sources which are artifacts at the edges in the column density maps. The derived clumps from each cloud are identified in Fig. \ref{fig:pol-plk-gal} using ellipses in white.

Fig. \ref{fig:hist_clumot} shows the distribution of the offsets derived from \textit{Astrodendro} method for the whole sample without classification. The correlation of the clump orientation with respect to the magnetic field for all the four clouds is shown in upper panel and the clump orientation with respect to the direction of motion is shown in lower panel. In L1147/1158, the distribution of $\Delta\theta^{clump}_{BICMF}$ and $\Delta\theta^{clump}_{Bcloud}$ is bimodal although it has preferential alignment of being perpendicular with more clumps at $>$60\degr. In L1172/1174, L1228 and L1251, the offsets are random. 
In L1147/1158, the distribution of the offsets of the clump major axis with respect to the motion of the cloud show more clumps in the range 0\degr-30\degr. 
Similarly, in L1172, the distribution of $\Delta\theta^{clump}_{motion}$ shows marginally a bimodal trend. In L1228, the offsets in the range $>$60\degr has maximum number of clumps suggesting that the clumps tend to be more perpendicular to the direction of motion. In cometary shaped cloud L1251, the offset distribution shows a bimodal distribution with either parallel or perpendicular orientations. There are 130 clumps lying $<$30\degr, 92 clumps lying in the range 30\degr$<\Delta\theta^{clump}_{motion}\leq$60\degr and 82 clumps having offsets higher than $>$60\degr. 

Having derived the histograms of the offsets for the elongated clumps using two different source extraction methods, we checked for the common sources. By putting a condition between the centroid positions of the clumps as three times the beam width, we checked for the matching sources from both the methods. The percentage of common sources for both the methods within this condition has been found to be 70$\%$, 80$\%$, 86$\%$ and 88$\%$ for L1147/1158, L1172/1174, L1228 and L1251, respectively. We find that the correlations of the core orientation with the magnetic field and the direction of motion does not follow any systematic trend with both the methods. The random distribution of the offsets of major axis with respect to magnetic field and the direction of motion in all the four clouds suggest that the core orientation depends on the individual cloud magnetization and the local dynamics. It is important to keep in mind that the analysis presented here were carried out on 2D projected maps and therefore the projection from 3D to 2D plane also can effect the measured core morphology. The analysis presented by various authors \citep{2020MNRAS.494.1971C,2014ApJ...791...43P} have also suggested that the relative orientation of the cores with the magnetic field is random in nearby star-forming regions like Lupus I, Taurus, Ophiuchus and Perseus clouds.


\section{Conclusions}\label{sec:con}

We present results of a study conducted on four molecular cloud complexes situated in the Cepheus Flare region, L1147/1158, L1172/1174, L1228 and L1251. Using the \textit{Gaia} EDR3 data for the YSOs associated with these clouds we estimated a distance of 371$\pm$22 pc and 340$\pm7$ pc for L1228 and L1251 respectively implying that all the four complexes are located at similar distances from us. Using proper motions of the YSOs with the assumption that both clouds and the YSOs are kinematically coupled, we estimated the projected direction of motion of the clouds. The clouds are found to be in motion at an offset of $\sim30$\degr~with respect to the ambient magnetic fields inferred from the \textit{Planck} polarization measurements except in the case of L1172/1174 in which the offset is $\sim45$\degr. The inner and outer magnetic field orientations are found to be nearly parallel suggesting that the cloud magnetic fields are inherited from the ICMF. 

In L1147/1158, the major axis of the starless clumps are found to be oriented predominantly parallel to both ICMF/cloud magnetic fields while it is around 40\degr~for prestellar cores. In L1172/1174, the major axis of the clumps are found to be aligned more random to the field lines for both starless and prestellar clumps. In L1228 and L1251 the offset between the major axis of the clumps and the field lines are found to be more random for starless clumps whereas the offsets are overall more around 40\degr-80\degr for prestellar clumps. It may be possible that the different preferred alignments or the random distribution of the cores' major axis may be related to the local magnetized properties of each cloud.

With respect to the motion of the clouds, there is a marginal trend of the prestellar clumps' major axis being more parallel in L1251 and bimodal distribution for starless clumps. For L1158, starless clumps are more oriented parallel but random distribution for prestellar clumps. In L1172/1174, the major axis for starless cores are oriented more randomly whereas for prestellar clouds, major axis follows roughly a bimodal distribution. In L1228, both the starless and prestellar cores show random distribution. The preferred alignment of core orientation with respect to the direction of motion which suggests that the projected direction of motion could be a regulator of the core dynamics.


\begin{acknowledgements}

This research has made use of data from the Herschel Gould Belt survey project (http://gouldbelt-herschel.cea.fr). The HGBS is a Herschel Key Project jointly carried out by SPIRE Specialist Astronomy Group 3 (SAG3), scientists of several institutes in the PACS Consortium (CEA Saclay, INAF-IAPS Rome and INAF-Arcetri, KU Leuven, MPIA Heidelberg), and scientists of the Herschel Science Center (HSC). This work has made use of data from the following sources: (1) European Space Agency (ESA) mission {\it Gaia} (\url{https://www.cosmos.esa.int/gaia}), processed by the {\it Gaia}
Data Processing and Analysis Consortium (DPAC,
\url{https://www.cosmos.esa.int/web/gaia/dpac/consortium}). Funding for the DPAC
has been provided by national institutions, in particular the institutions
participating in the {\it Gaia} Multilateral Agreement.; (2) the Planck Legacy Archive (PLA) contains all public products originating from the Planck mission, and we take the opportunity to thank ESA/Planck and the Planck collaboration for the same;  (3) the \textit {Herschel} SPIRE images from \textit {Herschel} Science Archive (HSA). \textit{Herschel} is an ESA space observatory with science instruments provided by European-led Principal Investigator consortia and with important participation from NASA. Some of the results in this paper have been derived using the healpy and HEALPix package. We also used data provided by the SkyView which is developed with generous support from the NASA AISR and ADP programs (P.I. Thomas A. McGlynn) under the auspices of the High Energy Astrophysics Science Archive Research Center (HEASARC) at the NASA/ GSFC Astrophysics Science Division.

\end{acknowledgements}

\bibliographystyle{aa} 
\bibliography{reference} 

\end{document}